\documentclass[sigconf]{acmart}
\AtBeginDocument{%
  \providecommand\BibTeX{{%
    \normalfont B\kern-0.5em{\scshape i\kern-0.25em b}\kern-0.8em\TeX}}}

\setcopyright{acmlicensed}
\copyrightyear{2024}
\acmYear{2024}
\acmDOI{10.1145/3658644.3690331}

\acmConference[ACM CCS]{ACM Conference on Computer and Communications Security}{October 14-18, 2024}{Salt Lake City, U.S.A.}
\acmISBN{979-8-4007-0636-3/24/10}

\usepackage{amsmath,amsfonts,pifont}
\usepackage{algorithmic}
\usepackage{graphicx}
\usepackage{textcomp}
\usepackage{xcolor}
\usepackage{subfiles} 
\usepackage{hyperref}

\usepackage{todonotes}

\usepackage{xcolor}
\newcommand{\getsr}{\ensuremath{\leftarrow_\$}}

\newcommand{\bits}[1]{\{0,1\}^{#1}}
\newcommand{\secpar}{\lambda}
\mathchardef\mhyphen="2D

\newtheorem{definition}{\textbf{Definition}}
\newtheorem{remark}{\textbf{Remark}}

\newcommand{\UE}{\mathrm{UE}}
\newcommand{\gNB}{\mathrm{gNB}}
\newcommand{\CN}{\mathrm{CN}}

\newcommand{\Alg}[1]{\ensuremath{\mathsf{#1}}}
\newcommand{\SanSig}{\Alg{SanSig}}

\newcommand{\KeyGen}{\Alg{KGen}}
\newcommand{\Sign}{\Alg{Sign}}
\newcommand{\Sanit}{\Alg{Sanit}}
\newcommand{\Verify}{\Alg{Verify}}

\newcommand{\Enc}{\Alg{Enc}}
\newcommand{\Dec}{\Alg{Dec}}

\newcommand{\Prove}{\Alg{P}}
\newcommand{\ZK}{\Alg{ZK}}
\newcommand{\com}{\Alg{Com}}

\newcommand{\adv}{\ensuremath{\mathcal{A}}}

\newcommand{\pk}{\ensuremath{\mathit{pk}}}
\newcommand{\sk}{\ensuremath{\mathit{sk}}}
\newcommand{\PK}{\ensuremath{\mathit{PK}}}
\newcommand{\SK}{\ensuremath{\mathit{SK}}}
\newcommand{\adm}{\ensuremath{\mathit{ADM}}}
\newcommand{\modd}{\ensuremath{\mathit{MOD}}}
\newcommand{\cert}{\ensuremath{\mathbb{C}}}
\newcommand{\pid}{\ensuremath{\mathit{pid}}}
\newcommand{\gid}{\ensuremath{\mathit{id_{gNB}}}}

\newcommand{\UID}{\mathrm{UID}}

\newcommand{\ck}{\ensuremath{\mathit{ck}}}

\newcommand{\ClientAction}[1]{ 
	\node[right] at (\InitX, \Y) {#1};
}
\newcommand{\ServerAction}[1]{
	\node[left] at (\RespX, \Y) {#1};
}

\newcommand{\AdversaryAction}[1]{
	\node at ($1/2*(\InitX, \Y)+1/2*(\RespX, \Y)$) {#1};
}
\newcommand{\ClientToServer}[3][->]{
	\NextLine[0.5]
	\draw[#1] (\ArrowLeft,\Y) -- node[above] {#2} node[below] {#3} (\ArrowRight,\Y) ;
	\NextLine[0.5]
}
\newcommand{\ServerToClient}[3][->]{
	\NextLine[0.5]
	\draw[#1] (\ArrowRight,\Y) -- node[above] {#2} node[below] {#3} (\ArrowLeft,\Y) ;
	\NextLine[0.5]
}
\newcommand{\ClientToAdversary}[3][->]{
	\NextLine[0.5]
	\draw[#1] (\ArrowLeft,\Y) -- node[above] {#2} node[below] {#3} (\ArrowCenter,\Y) ;
	\NextLine[0.5]
}

\newcommand{\AdversaryToClient}[3][->]{
	\NextLine[0.5]
	\draw[#1] (\ArrowCenter,\Y) -- node[above] {#2} node[below] {#3} (\ArrowLeft,\Y) ;
	\NextLine[0.5]
}
\newcommand{\AdversaryToServer}[3][->]{
	\NextLine[0.5]
	\draw[#1] (\ArrowCenter,\Y) -- node[above] {#2} node[below] {#3} (\ArrowRight,\Y) ;
	\NextLine[0.5]
}

\newcommand{\NextLine}[1][1.0]{
	\pgfmathparse{\Y+#1}
	\edef\Y{\pgfmathresult}
}
\newcommand{\Separatorb}[2][1.0]{
	\draw[very thick,black] (\InitX,\Y+0.5) node[above=-0.1cm,anchor=north west] {\bf #2} -- (\RespX,\Y+0.5);
}

\newcommand{\linkgame}[2]{\hyperref[#1]{G#2}}

\newcounter{Bdversary}

\newcommand\orcidicon[1]{\href{https://orcid.org/#1}{\mbox{\scalerel*{
\begin{tikzpicture}[yscale=-1,transform shape]
\pic{orcidlogo};
\end{tikzpicture}
}{|}}}}

\newcommand*{\addFileDependency}[1]{
  \typeout{(#1)}

  \IfFileExists{#1}{}{\typeout{No file #1.}}
}

\newcommand{\cmark}{\ding{51}}%
\newcommand{\xmark}{\ding{55}}
\usepackage{comment}
\usepackage{adjustbox}
\usepackage{tikz}
\usetikzlibrary{svg.path}
\usetikzlibrary{calc}
\usepackage{scalerel}
\usepackage{color}
\usepackage{soul,xcolor}
\usepackage{makecell}
\usepackage{enumitem}
\usepackage{mathtools}
\usepackage{newtxmath}
\usepackage{comment}
\usepackage{multirow}
\usepackage{url}
\usepackage{wrapfig}
\usepackage{diagbox}
\usepackage{subcaption}
\usepackage{cleveref}

\newif\iffullversion
\newif\ifsubmissionversion
\fullversionfalse
\submissionversiontrue

\usepackage{relsize}
\usepackage{booktabs}

\begin{document}


\title [Strong Privacy-Preserving UC AKA with Handover for MVNO]{Strong Privacy-Preserving Universally Composable \\AKA Protocol with  Seamless Handover Support for \\ Mobile Virtual Network Operator}


\author{Rabiah Alnashwan}
\authornote{Both authors contributed equally to this research.}
\affiliation{%
\institution{The University of Sheffield}
\city{}
    \country{}}
\email{ralnashwan1@sheffield.ac.uk}
\affiliation{%
\institution{Imam Mohammad Ibn Saud Islamic University}
\city{}
    \country{}}
\email{ralnashwan@imamu.edu.sa}

\author{Yang Yang}
\authornotemark[1]
\authornote{Corresponding Author.}
\affiliation{%
 \institution{National University of Singapore}
    \city{}
    \country{}
}
\email{y.yang@u.nus.edu}

\author{Yilu Dong}
\affiliation{%
 \institution{The Pennsylvania State University}
 \city{}
    \country{}
}
\email{yiludong@psu.edu}

\author{Prosanta Gope}
\authornotemark[2]
\affiliation{%
 \institution{The University of Sheffield}
 \city{}
    \country{}
}
\email{p.gope@sheffield.ac.uk}

\author{Behzad Abdolmaleki}
\affiliation{%
 \institution{The University of Sheffield}
 \city{}
    \country{}
}
\email{behzad.abdolmaleki@sheffield.ac.uk}

\author{Syed Rafiul Hussain}
\affiliation{%
 \institution{The Pennsylvania State University}
 \city{}
    \country{}
}
\email{hussain1@psu.edu}


\renewcommand{\shortauthors}{}

\begin{abstract}
Consumers seeking a new mobile plan have many choices in the present mobile landscape. The Mobile Virtual Network Operator (MVNO) has recently gained considerable attention among these options. MVNOs offer various benefits, making them an appealing choice for a majority of consumers. These advantages encompass flexibility, access to cutting-edge technologies, enhanced coverage, superior customer service, and substantial cost savings.  Even though MVNO offers several advantages, it also creates some security and privacy concerns for the customer simultaneously. For instance, in the existing solution, MVNO needs to hand over all the sensitive details, including the users' identities and master secret keys of their customers, to a mobile operator (MNO) to validate the customers while offering any services. This allows MNOs to have unrestricted access to the MVNO subscribers' location and mobile data, including voice calls, SMS, and Internet, which the MNOs frequently sell to third parties (e.g., advertisement companies and surveillance agencies) for more profit. Although critical for mass users, such privacy loss has been historically ignored due to the lack of practical and privacy-preserving solutions for registration and handover procedures in cellular networks. In this paper, we propose a universally composable authentication and handover scheme with strong user privacy support, where each MVNO user can validate a mobile operator (MNO) and vice-versa without compromising user anonymity and unlinkability support.  Here, we anticipate that our proposed solution will most likely be deployed by the MVNO(s) to ensure enhanced privacy support to their customer(s).  

\end{abstract}



\maketitle

\section{Introduction} 
\label{sec:intro}

\begin{tikzpicture}[remember picture, overlay]
    \node at (-0.5,15.5) {This is the full version of the paper that has been accepted by \textit{ACM Conference on
Computer and Communications Security} (ACM CCS 2024).};
\end{tikzpicture}

Mobile Virtual Network Operators (MVNOs) are making significant strides in telecommunications by capitalizing on existing wireless network infrastructure \cite{li2020understanding, vallina2015beyond}. Through acquiring network capacity from Mobile Network Operators (MNOs) like Vodafone and T-Mobile, MVNOs such as Virgin Mobile and Google Fi have swiftly carved out a niche in the market, employing a business model focused on delivering cost savings to consumers. This strategic framework not only allows MVNOs to present more budget-friendly pricing options but also positions them as compelling alternatives to traditional MNOs. 

Now, with the advent of 5G, the fifth generation of wireless technology, MVNOs are empowered with faster speeds, lower latency, increased capacity, and the ability to support a wide range of connected devices. This has improved their existing services and opened up opportunities to explore new business models and expand their offerings \cite{kim2007economic, shin2007study, sacoto2020game}. This may include enhanced IoT connectivity, immersive multimedia experiences, and other applications that benefit from the capabilities of 5G. However, the inherent vulnerabilities of cellular network technology, such as the lack of authentication of broadcast messages\cite{rupprecht2018security, hussain2019insecure, singla_look_2021, lotto2023baron}, missing confidentiality and integrity protections for pre-authenticated and even for some post-authenticated messages \cite{shaik2016practical, park2022doltest, hussain2021noncompliance, yu2023secchecker, bitsikas2023ue, kim2023basecomp, hussain20195greasoner}. Therefore, it is imperative that MNOs and MVNOs work together to implement appropriate security measures and ensure the privacy and protection of user data.

\begin{figure}[]
    \centering
    \includegraphics[width=0.95\linewidth]{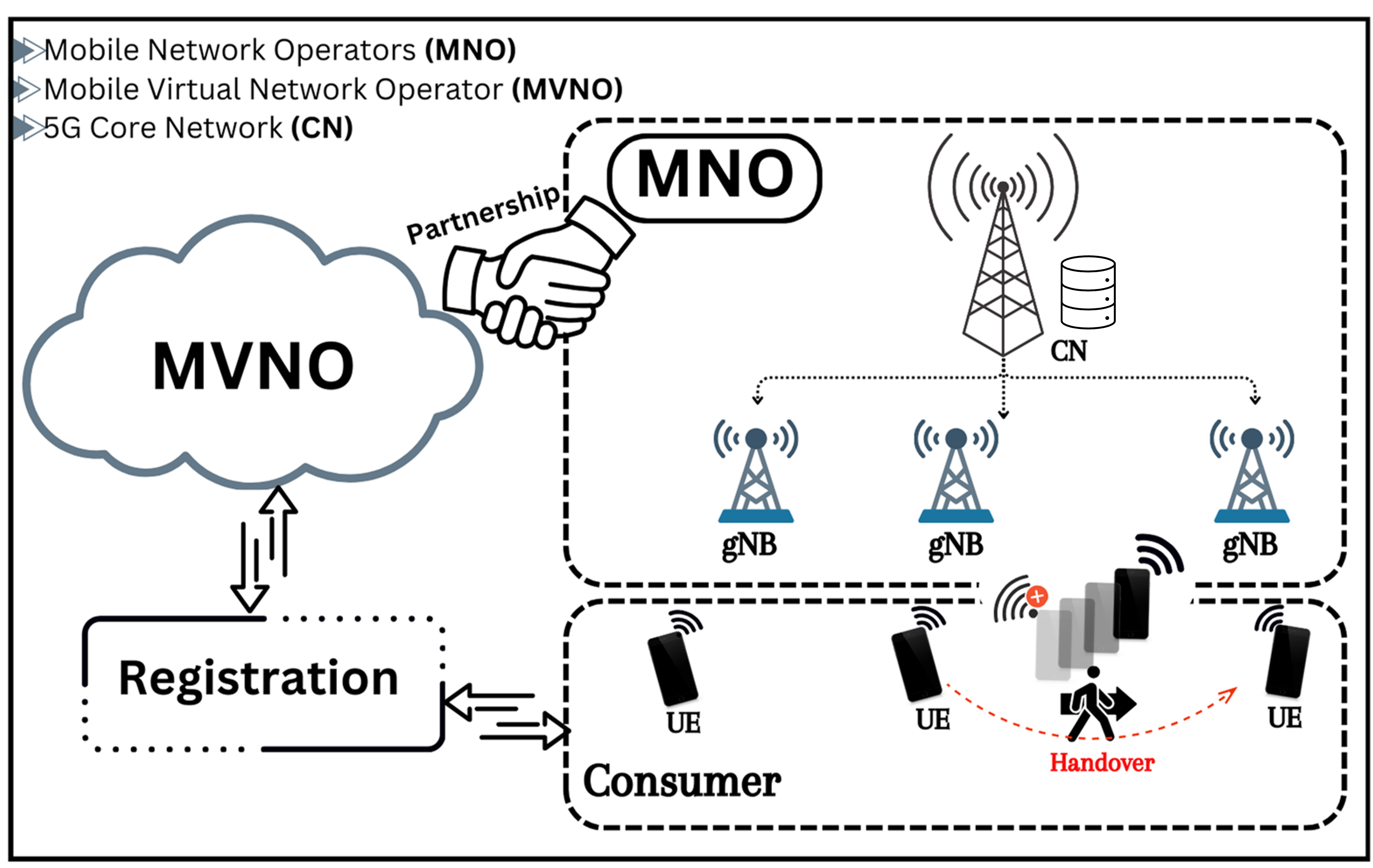}
    \vspace{-.3cm}
    \caption{Proposed System Model}
    \label{fig:System Model}
    \vspace{-5mm}
\end{figure}

The operational foundation of an MVNO (e.g., Virgin Mobile in the UK) hinges on the infrastructure of another operator (O2 in the UK) through a well-established partnership. Consequently, a user registered with an MVNO effectively utilizes the network infrastructure provided by MNO and seamlessly accesses MNO's infrastructure within this collaborative arrangement while maintaining registration with MVNO, the only operator they trust. This cooperation between MVNOs and established MNOs underscores the pivotal role of secure and privacy-preserving \textit{Authentication} and \textit{Handovers} in MVNO environments \footnote{Refer to Figure \ref{fig:System Model} for a clearer view of the scenario} and emphasizes the need to address the following key concerns in this dynamic domain.

\textit{\underline{P1: Preventing identity exposure:}} 
During both the Authentication and Key Agreement (AKA) and the handover procedures, users engage with MNO infrastructure (i.e., base stations and core network), which are considered third parties from the MVNO users' perspective. Consequently, users reveal their identities (e.g., Subscription Permanent Identifier or SUPI and other temporary identifiers) to these third parties, i.e., MNOs, for authentication and handover purposes. Thus, they break entirely the anonymity guarantees. 
This allows the third-party MNO  to access users' footprints and link their activities, making them susceptible to tracking and fingerprinting attacks \cite{hussain2018lteinspector, cremers_component-based_2019,basin_formal_2018}.
Moreover, the current 5G AKA protocol has been shown to be vulnerable to linkability attacks despite attempts to enhance user privacy. Recent research has demonstrated that an attacker can still link 5G AKA sessions and trace users' movements by exploiting protocol weaknesses, particularly in the failure messages \cite{borgaonkar_new_2018}. This vulnerability persists even with the introduction of concealed long-term identifiers, highlighting a critical gap in user privacy protection within the existing 5G infrastructure that extends to MVNO scenarios.

Unfortunately, this issue poses a significant privacy concern as MNOs consistently sell and expose the mobile location browsing data and call metadata of millions of users. This behaviour by MNOs leaves users with no recourse due to a deregulated industry, widespread mobile use, and the prevalence of data brokers. Hence, it becomes imperative to minimize the disclosure of user information or, ideally, eliminate any exposure. 
The current literature lacks a fully anonymous, unlinkable, and footprint-free protocol despite these privacy concerns. Such a protocol is essential to enable users to receive the required services without compromising their privacy.

\textit{\underline{P2: Privacy-preserving mutual authentication:}} MVNO users must mutually authenticate the MNOs and their base stations to confirm the authenticity of the network providers and prevent fake base station attacks without exposing their footprints. However, ensuring the authenticity of a third-party provider (MNO) with whom the user is not registered, and vice versa, presents a challenge that has been overlooked in existing literature. Addressing these aspects becomes paramount to establishing a comprehensive security framework in the evolving 5G and MVNO interactions landscape.

\textit{\underline{P3: Privacy-preserving and secure revocation:}} An additional challenge arises when considering the need for user revocation in these networks. User revocation, the process of terminating/deactivating a user's access to network services, is important for managing resources and security. Implementing revocation while preserving anonymity is particularly challenging in the complex MNO-MVNO environment, as the current 5G protocols~\cite{3rd_generation_partnership_project_3gpp_security_2020} require exposing user identities to MNOs during revocation, thereby compromising the privacy we aim to preserve. This introduces a new dimension to the privacy problem in MNO-MVNO networks that must be addressed alongside authentication and handover privacy concerns.

\textit{\underline{P4: Universal Composability:}} Third, a crucial property, especially when deployed in large systems like 5G and MVNO, is the ability to securely compose protocols to obtain more complex ones, as achieved through the Universal Composability (UC) framework \cite{canetti2001universally}. Relying on the UC framework enables arbitrary and secure protocol composition in a modular manner (e.g., compositions of AKA and handover), ensuring that the security guarantees of individual components extend to the entire system. This capability is vital for 5G systems where multiple cryptographic protocols must harmonize to provide robust security while preserving user privacy. Ongoing research and development efforts in academia and industry are dedicated to enhancing their composability and versatility to meet the ever-evolving demands of the digital age \cite{abdolmaleki2023universally,canetti2020universally,canetti2015simpler,gajek2008universally,abdolmaleki2020lift}. However, in the context of MVNOs, this security property has not been studied thus far, and the construction of UC-secure MVNOs remains unknown. In the context of MVNOs, which operate in dynamic and complex network environments, universally composable security offers several advantages, including (a) \textit{comprehensive security analysis}, i.e., identifying and mitigating threats by considering various possible interactions with other components and protocols, (b)  \textit{interoperability}, i.e, ensuring security guarantees even in the presence of the interactions with MNOs and users, (c) \textit{flexibility and adaptability}, in other words, allowing MVNOs to adapt and evolve their security measures without sacrificing the overall security posture even when the mobile network environment is subject to changes and advancements, and (d) \textit{formal verification} which allows for mathematical proofs of security properties.

In the pursuit of addressing the above concerns and enhancing security and privacy within 5G-enabled MVNO environments, we make the following contributions. 

\begin{itemize}[leftmargin=0.3cm]

    \item \emph{We design privacy-preserving AKA and revocation mechanisms} for an MVNO environment that supports a new notion of practical ZKP that aligns with the revised definition of revocation. This definition maintains user privacy by utilizing non-identifying yet unique attributes to facilitate revocation, ensuring legal compliance without disclosing user identities. We then present a novel approach (using a list of hash of identity of the user and designing ZKP for that) to design two security protocols that integrate ZKP with Universal Composability (UC) security. To the best of our knowledge, this is the \textit{first} solution to offer a universal solution to address the challenges linked to secure and privacy-preserving schemes within the MVNO environment.
  
    \item We develop a \emph{secure privacy-preserving handover protocol} that provides seamless user handover without central entity intervention, reducing overhead on the core network;
    
    \item We introduce a \textit{new} notion of user privacy, i.e., \emph{"Comprehensive Privacy"} in the context of an MVNO environment. This concept guarantees the anonymity of the user within the network, ensuring that the user's identity remains protected not only from the core network but also from all base stations controlled by MNO. Besides, our proposed scheme ensures security against linkability attacks and fake base station attacks.
    
    \item We perform a rigorous formal analysis of the security and privacy properties of our proposed schemes based on a comprehensive \emph{Universal Composability} framework.

    \item We implemented and evaluated our proposed schemes in an open-source 5G testbed. Our schemes achieve the desired security and privacy guarantees with only 0.19s overhead compared to conventional AKA. We also compared our approach with existing works aiming to enhance 5G AKA. Our approach can provide stronger security properties while increasing less overhead.
    We release the source code of our schemes and the existing approaches to which we compare at \cite{github_repo}.
\end{itemize}

Finally, we give our notation used in this paper in Table \ref{tab:Notation}.

\begin{table}[b]
 \caption{Notation used in our proposed scheme.}
 \vspace{-0.3cm}
 \label{tab:Notation}
\begin{tabular}{ll}
\hline
Notation & Description \\ \hline
$\PK_{u},\SK_{u}$  &     user's Public/secret keys for Enc/Dec   \\
$s\pk,s\sk$   &     Public/secret keys for signature   \\ 
$\PK^{g}_{san},\SK^{g}_{san}$    &     Public/secret keys for sanitising signatures   \\ 
 $\cert_{G}$      &     $\gNB$ certificates consists of ($\cert_{mod} $\&$ \cert_{fix}$)       \\ 
 $\sigma_{G}$ & $\gNB$ certificates' signature\\
 $\pid$ & user pseudo identity \\
 $\Enc/\Dec$  & Encryption and decryption            \\ 
$crs$& A common reference string\\
$ck$& A commitment key\\
  \hline
\end{tabular}
\end{table}

\section{Preliminaries} 
\label{sec: Preliminaries}

This section first provides a brief primer on MVNOs and then presents and defines the security of the cryptographic primitives that are fundamental to the construction of the proposed scheme. 

\subsection{Mobile Virtual Network Operator}

A Mobile Virtual Network Operator (MVNO) is a wireless communication service provider that operates without owning the physical infrastructure of the wireless network it utilizes to deliver services to its customers. Instead of deploying its own network infrastructure, an MVNO leases network capacity from a Mobile Network Operator (MNO). The MNO, which owns the network infrastructure, provides the necessary services, and the MVNO essentially acts as a reseller, offering mobile services under its brand. MVNOs are categorized into three types based on their dependence on host carriers: \emph{skinny}, \emph{light}, and \emph{thick}. Skinny MVNOs heavily rely on base carriers, with limited control over network elements, focusing on branding and marketing. Light MVNOs strike a balance between dependence and autonomy, offering additional services for differentiation. Thick MVNOs have the highest autonomy, controlling essential network elements, allowing flexibility in services, and sometimes investing in their infrastructure. These classifications illustrate the varying levels of control and differentiation within the mobile telecommunications ecosystem.
Among the three categories, Thick MVNOs are the least common. Therefore, this study focuses on MVNO types primarily relying on the MNO infrastructure for their services, such as Skinny and Light MVNOs.

\subsection{5G AKA and Handover}
\label{subsec:aka_handover}

\noindent \textbf{5G-AKA.}
In the existing 5G-AKA protocol, a base station (gNBs) periodically broadcasts its cell information to inform nearby phones, i.e., User Equipment (UEs) of the gNB's presence. Although these broadcast messages allow the UEs to connect to the gNB,  5G-AKA does not provide any mechanism for UEs to authenticate these initial broadcast messages. This leads to the fake base station attacks~\cite{hussain20195greasoner, hussain2019insecure}. After the initial connection between a UE and a gNB, the 5G-AKA protocol provides mutual authentication between the User Equipment (UE) and the Core Network (CN) through a sequence of control-plane messages \cite{3rd_generation_partnership_project_3gpp_security_2020}. Session keys are also derived to maintain secure communication between UE, gNB, and CN. However, the current mutual authentication scheme requires the UE to send its identity (SUPI) over-the-air (OTA), before the shared session key is derived. In this case, an attacker can sniff the OTA packets and learn the sensitive user data, e.g., SUPI (if unencrypted) and temporary identifiers, which allow a Dolev-Yao attacker~\cite{dolev_security_1983} to track the user. Although the user identity can be sent in an encrypted form (SUCI), the MNO usually needs to decrypt the user identity to authenticate the user and provide the service. The MNO can also decrypt user plane traffic, enabling them to monitor and analyze user Internet activity. This allows the MNO to gain insights into the types of websites visited, the applications used, and overall user behaviour online. Nonetheless, the inherent flaws in the 5G AKA protocol, such as lack of observational equivalence in certain sub-protocols, lead to the linkability/traceability ttacks~\cite{hussain20195greasoner, basin_formal_2018}. 

\noindent \textbf{Handover.} Handover is essential in cellular communication to provide seamless connections \cite{alnashwan_privacy-aware_2023}. Since the base stations are stationary, mobile devices need to switch the connected base stations frequently. Here, we discuss \textit{three distinguished types of handovers}: intra-cell handover, inter-cell handover, and inter-MNO handover.

\textit{Intra-cell handover.} In intra-cell handover, the UE needs to disconnect the current radio connection and establish a new connection (usually in a different frequency) within the same cell. During this type of handover, an attacker can use a fake base station to impersonate the cell-to-be-connected and try to get the user's identity. 

\textit{Inter-cell handover.} In inter-cell handover, the UE seamlessly moves from one base station (source) to another (target). It needs to re-establish the radio connection with the target base station. In this case, UE is also vulnerable to fake base station attacks. In addition, in inter-cell handover, the source base station sends the UE context (e.g., UE's and base station's identities, UE security context, and PDU session information) to the target gNB either through direct communication, i.e., using Xn interface between two gNBs or through the MNO's CN, i.e, with N2 interface between gNBs and CN. As a result, the infrastructure owners (MNOs), is able to track user movements from the handover requests.

\textit{Inter-MNO handover.} Inter-MNO handover is specific to the MVNO settings. An MVNO user can use the base station infrastructure from several MNOs to provide better radio coverage. To provide seamless handover, an interface between AMFs of different MNOs must be implemented to transfer the UE context. The MNOs have information about when the user enters and leaves its network and also which MNO it moves to. However, the MNOs usually are not motivated to implement this interface. Hence, UE must de-register from the current MNO and register with the new MNO, which causes an interruption of the current service.

\subsection{Sanitizable signatures}

Sanitizable Signature is a signature scheme where signing capabilities can be delegated to another party: the sanitiser. The accuracy of the sanitiser capabilities can be managed through a couple of deterministic functions $\adm$ and $\modd$.  The former indicates specific parts of which a sanitiser can modify a message. The latter function specifies what modification the sanitiser has made, where $m^{*}=\modd (m)$ and $\adm (m^{*},m)\to \{0,1\}$. A $\SanSig$ is a tuple of algorithms: $\SanSig$$\ = \ \{ \mathsf{Kgen}, \mathsf{Sign}, \mathsf{Sanit}, \mathsf{Verify}, \mathsf{Proof}, \mathsf{Judge} \}$

\begin{itemize}
    \item $\mathsf{Kgen}: (pk_{sig}, sk_{sig}) \leftarrow_\$ \mathsf{Kgen}(1^\lambda),\\ (pk_{san}, sk_{san}) \leftarrow_\$ \mathsf{Kgen}_{san}(1^\lambda)  $
    \item $\mathsf{Sign}: \sigma \leftarrow_\$ \mathsf{Sign}(m, sk_{sig}, pk_{san}, ADM) $
    \item $\mathsf{Sanit}:(m^*, \sigma^*)\leftarrow_\$ \mathsf{Sanit}(m, \modd, \sigma, pk_{sig}, sk_{san})  $
    \item $\mathsf{Verify}: b \leftarrow \mathsf{Verify}(m, \sigma, pk_{sig}, pk_{san}) $
\end{itemize}

Sanitizable Signature Unforgeability:
We say that $\SanSig$ supports Existential Unforgeability under Chosen Message Attack (EUFCMA-Secure) if the advantage of $Adv_{SanSig}(\mathcal{A})$ is negligible, where:
\begin{center}

    $Adv^{EUC-CMA}_{SanSig}(\mathcal{A}) = Pr[Exp^{EUC-CMA}_{SanSig}(\mathcal{A})=1]$
    
\end{center}

\subsection{Universal Composability}
  \newcommand{\inp}{\mathsf{x}}
  \newcommand{\wit}{\mathsf{w}}
   \newcommand{\crs}{\mathsf{crs}}
   \newcommand{\prover}{\mathsf{P}}
   \newcommand{\kcrs}{\mathsf{K}_{\crs}}
   \newcommand{\verifier}{\mathsf{V}}
   \newcommand{\simulator}{\mathsf{Sim}}
   \newcommand{\LANG}{\mathcal{L}}
\newcommand{\envir} {\mathcal{Z}}
The Universal Composability (UC) framework was introduced by Canetti in \cite{canetti2001universally}. In the UC framework, one analyzes the protocol's security under real-world and ideal-world paradigms. More precisely, in this setting, the real-world execution of a protocol is compared with an ideal-world interaction with the primitive it implements.
Then, a composition theorem in this model states that the security of the UC-secure protocols remains if it is arbitrarily composed with other UC-secure protocols or the protocol itself.
Additionally, the UC-secure property guarantees security in practical applications where individual instances of protocols are run in parallel, such as the Internet.
The entities in the UC framework in both ideal-word and real-word executions are modelled as PPT (probabilistic polynomial time) interactive Turing machines that send and receive messages through their output and input tapes, respectively. 
In the ideal world execution, dummy parties (possibly controlled by an ideal-word adversary $\simulator$, also called simulator) communicate directly with the ideal functionality $\mathcal{F}$. The ideal functionality can be viewed as a trusted party that creates the primitives to implement the protocol.
Correspondingly, in the real-world execution,  parties (possibly corrupted by a real-world adversary $\adv$) communicate with each other as a protocol $\Pi$ that realizes the ideal functionality. 
Both the ideal and real executions are controlled by the environment $\envir$, an entity that sends inputs and receives the outputs of $\adv$, the individual parties, and $\simulator$. 
Finally, after seeing the ideal or real protocol execution, $\envir$ returns a bit, which is considered the execution output. 
Then, the rationale behind this framework lies in showing that the environment $\envir$ can not efficiently distinguish between the ideal and real executions, therefore meaning that the real-world protocol is as secure as the ideal-world (the ideal functionality).

Besides the two aforementioned models (real-world and ideal-world) of computation, the UC framework considers the hybrid world, where the executions are similar to the real world but with the additional assumption that the parties are allowed to access an auxiliary ideal functionality $\mathcal{G}$.
More precisely, in this case, instead of honest parties interacting directly with the ideal functionality, the adversary passes all the messages from and to the ideal functionality. 
Also, the transmission channels are considered to be ideally authenticated, meaning that the adversary is not able to modify the messages but is only able to read them.  
Unlike information transferred between parties, which can be read by the adversary, the information transferred between parties and the ideal functionality is split into a public and private header. 
The private header carries some information like as the private inputs of parties and it cannot be read by the adversary. 
The public header carries only some information that can be viewed publicly, such as receiver,  sender, type of message,  and session identifiers.
Let denote the output of the environment $\envir$ that shows
the execution of a protocol $\Pi$ in a real-world model and a hybrid model, respectively, as $\textsc{Ideal}^{\mathcal{F}}_{\simulator}$ and $\textsc{Hybrid}^{\mathcal{G}}_{\Pi, \adv}$. Then the UC security is formally defined as:
\begin{definition}
	A n-party ($n \in \mathbb{N}$) protocol $\Pi$  UC-realizes an ideal functionality $\mathcal{F}$ in the hybrid model if, for every PPT adversary $\adv$, there exists a simulator $\simulator$ such that for all environments $\envir$,
	\[
	\textsc{Ideal}^{\mathcal{F}}_{\simulator} \approx_\secpar \textsc{Hybrid}^{\mathcal{G}}_{\Pi, \adv}.
	\] 
	The protocol $\Pi$ is statistically secure if the above definition holds for all unbounded $\envir$. In \cref{thesis-sec:UC}, we define the ideal functionality for a commitment scheme and provide its corresponding hybrid functionality to prove the UC security of the scheme.
\end{definition}

\subsection{Commitment Scheme}
\begin{definition}
	A commitment scheme $\Pi = (\mathsf{Kgen}, \mathsf{Com}, \mathsf{Decom)}$, is defined by the following three algorithms:
	\begin{itemize}[leftmargin=0.6cm]
		\item $\mathsf{ck} \gets \mathsf{Kgen} (\lambda)$: given a security parameter $\lambda$, generates a public parameter $\mathsf{ck}$ of the scheme that implicitly
		passed as input to the other algorithms.
		\item $(\mathsf{c}, \mathsf{\delta}) \gets \mathsf{Com}_{\mathsf{ck}} (m, r)$: given the public parameter $\mathsf{ck}$, a message $m$ from message space $M$, and a randomness $r$ from randomness space $R$, outputs a commitment $\mathsf{c}$ together with an opening information $\mathsf{\delta}$\footnote{\textbf{The opening information will be used in the decommit phase to prove that the commitment $\vec{c}$ contains a valid message $m$.}}.
		\item $m / \bot \gets \mathsf{Decom} (\mathsf{ck}, \mathsf{c}, m, \mathsf{\delta})$: given given the public parameter $\mathsf{ck}$, a commitment $\mathsf{c}$, the message $m$ and an opening information $\mathsf{\delta}$, outputs $m$ or $\bot$ if the opening verification fails.
	\end{itemize}
\end{definition}
Such a scheme must satisfy both \textit{hiding} property (meaning that the commit phase does not disclose any information about the committed message $m$), and \textit{binding} property (meaning that the decommit phase (opening phase) can successfully open to only one value).
The aforementioned properties may be achieved in a perfect, statistical or computational according to the power of the adversary against those properties.
Besides, some additional strong properties are demanded in some systems, like the UC-secure commitment scheme. The first is \textit{extractability}, which states that given a trapdoor, one (i.e., the simulator $\mathsf{Sim}$ in the UC model) can recover the committed value $m$. The second one is \textit{equivocability}, which means that given a trapdoor, one (i.e., the simulator $\mathsf{Sim}$ in the UC model) can open a commitment to any message $m'\neq m$.

\subsection{Non-Interactive Zero-Knowledge} 
Zero-knowledge proofs and in particular, non-interactive zero-knowledge proofs (NIZKs), is a protocol between a prover and a verifier that allows the prover to convince the verifier of the validity of a statement without disclosing any more additional information. 
Let $\mathsf{REL}$ be a relation generator, such that $\mathsf{REL} (1^\lambda)$ outputs a polynomial time decidable binary relation $\mathcal{R} = \{(\inp, \wit)\}$.
Here, $\inp$ and $\wit$ are, respectively, the statement and the witness.
Let $\LANG_{\mathcal{R}} = \{\inp: \exists \wit, (\inp, \wit) \in \mathcal{R} \}$ be an $\sf NP$-language.
NIZK proofs in the CRS model consist of the four algorithms $(\kcrs, \prover, \verifier, \simulator)$ where $\kcrs$, $\prover$, $\verifier$, and $\simulator$ are common reference strings (CRS) generator, prover, verifier, and the simulator, respectively.  
\newcommand{\tc}{\mathsf{td}}
\begin{definition}
   A NIZK system $\Psi$ for any relation generator $\mathsf{REE}$ consists of four PPT algorithms:
   \begin{itemize} [leftmargin=0.6cm]
	\item $(\crs, \tc) \gets \kcrs(\lambda)$: A \emph{probabilistic} algorithm that, given the security parameter $\lambda$ outputs a \emph{CRS trapdoor} $\tc$ and a CRS $\crs$. Otherwise, it outputs $\bot$.
	\item $\pi \gets \prover(\crs, \inp, \wit)$: A probabilistic algorithm that, given $(\crs, \inp, \wit)$, outputs an argument $\pi$ if $(\inp, \wit) \in \mathcal{R}$.
	Otherwise, it outputs $\bot$.
	\item $0/1 \gets \verifier(\crs, \inp, \pi)$: a probabilistic algorithm that, given $(\crs, \inp, \pi)$, returns either $0$ (reject) or $1$ (accept). 
	\item $ \pi \gets \simulator(\crs, \tc, \inp)$:] a probabilistic algorithm that, given $(\crs, \tc, \inp)$ outputs an argument $\pi$.
	Otherwise, it outputs $\bot$.
 \end{itemize}
\end{definition}

A NIZK must provide the following properties:
\begin{enumerate}
	\item \textbf{Completeness.}
	For any $\lambda$, and $(\inp, \wit) \in \mathcal{R}$,
	\[
	\Pr\left[
	\begin{aligned}
	&
	(\crs, \tc) \gets \kcrs(\lambda):
	\verifier (\crs, \inp, \prover (\crs, \inp, \wit)) = 1
	\end{aligned}
	\right] = 1
	\enspace.
	\]
	\item \textbf{Statistical Zero-Knowledge.} 
	For any computationally unbounded adversary $\adv$, $ |\varepsilon^{zk}_0 - \varepsilon^{zk}_1| \approx_\lambda0$, where $\varepsilon^{zk}_b :=$
	\[
	\mathsf{Pr}\left[
	\begin{aligned}
	&
	(\crs, \tc) \gets \kcrs(\lambda), b \gets \{0, 1\}: 
	\adv^{\mathsf{O}_b (\cdot, \cdot)} (\crs) = 1
	\end{aligned}
	\right].
	\]
	
	The oracle $\mathsf{O}_0 (\inp, \wit)$ returns $\bot$ (reject) if $(\inp, \wit) \not\in \mathcal{R}$, and otherwise it returns $\prover (\crs, \inp, \wit)$.
	Similarly, $\mathsf{O}_1 (\inp, \wit)$ returns $\bot$ (reject) if $(\inp, \wit) \not\in \mathcal{R}$, and otherwise it returns $\simulator (\crs, \tc, \inp)$.
	\item \textbf{Computational Soundness.}
	For any non-uniform PPT $\adv$,
	\[
	\mathsf{Pr}\left[
	\begin{aligned}
	&
	(\crs, \tc) \gets \kcrs(\lambda); 
	(\inp, \pi) \gets \adv (\crs):
	\\ &
	\verifier (\crs, \inp, \pi) = 1 \land 
	\neg (\exists \wit: (\inp, \wit) \in \mathcal{R} )
	\end{aligned}
	\right]
	\approx_\lambda 0
	\enspace.
	\]
\end{enumerate}

\vspace{4mm}

\section{System and Threat Model}
This section presents the system model and the threat model we consider in this work.

\subsection{Proposed System model}
The system model is outlined into three main layers: the \emph{MVNO}, \emph{MNO,} and \emph{Consumer} layers, as illustrated in Figure \ref{fig:System Model}. The MVNO is critical in generating essential security parameters, encompassing keys and IDs, for its users, namely UE. The MVNO typically does not own any mobile network infrastructure and rents the services leveraging MNO infrastructure such as Virgin, Lebara, and Tesco. The second layer of the system model comprises two essential participants--- the Core Network (CN) and the base station (gNB). Both entities in this layer share mutual trust, usually owned by a traditional mobile network operator (MNO), meaning they own the infrastructure they use, such as O2 and Vodafone. However, it is important to note that despite being owned by the same company, no secure communication channel is assumed between entities within this layer. The final layer in the proposed system model is the consumer layer, consisting of MVNO's users or user equipment; these terms are used interchangeably to denote the same entity.

\subsection{Threat Model}
Our work focuses on scenarios where the network entities, such as CN and gNBs, are controlled by an MNO, and all connected gNBs belong to this same MNO. We specify a threat model that includes a typical protocol-level Dolev-Yao adversary, an adversary against the privacy of the UE, and an adversary against the UC model. 
For this, we separate each adversary into one of \emph{three types} and define our threat model as follows:

    \noindent$\bullet$ [$\adv_1$] The Type 1 (Dolev-Yao) adversary consists of the typical Dolev-Yao model~\cite{dolev_security_1983}, which is capable of eavesdropping on the network between the UE and the entities operated by MNO (e.g., gNB, CN). This type of adversary can also use fake base station~\cite{hussain2019insecure} or machine-in-the-middle (MitM) relay~\cite{rupprecht2018security} to add, drop or modify messages between UE and MNO by adhering to the cryptographic assumptions, i.e., it can decrypt an encrypted message only if she obtains the decryption key.
    
    \noindent$\bullet$ [$\adv_2$] The Type 2 (Privacy) adversary is intent on compromising user privacy by attempting to breach the user's anonymity and establish links between the user's activities. This is broadly known as  \emph{unlinkability} or \emph{observational equivalence}. In the MVNO context, MNO's gNB and CN Type 2 adversaries as they are interested in users' footprint. Although these adversaries may not be able to compromise the network (MNO) entities. Still, it maliciously intends to collect and sell users' sensitive information to third parties. In addition, a Dolev-Yao adversary, i.e., $\adv_1$ may also attempt to track users' footprint by violating the observational equivalence property.

\noindent$\bullet$ [$\adv_3$] Finally, we consider a Type 3 (UC-based) adversary, which is more powerful and can run both the $\adv_1$ and $\adv_2$ internally. 

Unlike a Dolev-Yao adversary ($\adv_1$), which cannot reason about the security and privacy of multiple and unbounded parallel executions of the protocol, the $\adv_3$ can see many executions of the protocols when they are composed of other systems. Here, by having many executions in parallel, the adversary can verify observational equivalence properties that may find linkability/traceability attacks~\cite{hussain20195greasoner, basin_formal_2018}. 
The $\adv_3$ adversary can also infer and control/manipulate the internal protocol states of the victim. For instance, MNO's CN can infer the internal protocol states of the UE and can control those states by initiating the AKA or other common procedures at any time. The adversary can operate in a setting with multiple concurrent protocol instances, interacting with all instances simultaneously.

Unlike $\adv_1$ adversary, which is limited to symbolic operations and does not break cryptographic primitives, the UC adversary is more powerful, and can have full control over the network and adaptively and momentarily compromise parties at any time, obtaining their entire internal states. For instance, an external adversary may compromise a gNB by 
exploiting misconfigurations~\cite{LTE_misconfig}, vulnerabilities in implementation~\cite{LTE_misconfig}, dependency weaknesses~\cite{9222252}, and potential compromises by malicious user devices~\cite{CVE-2021-45462}.

To summarize, the attacker in the universally composable  
security model retains the Dolev-Yao capabilities and can perform the following actions in the context of MVNO: \textit{ (A) Initiating Protocol Sessions}: The adversary can initiate multiple protocol instances and interact with honest parties. \textit{(B) Choosing Instances of Sub-protocols}: The adversary can select instances of sub-protocols used within the larger protocol, potentially choosing those that benefit its malicious goals.
\textit{(C) Internal State Manipulation}: The adversary may attempt to manipulate the internal state of the protocol participants, influencing their behaviour. \textit{(D) Choosing Cryptographic Keys}: The adversary might have control over the generation or selection of cryptographic keys used in the protocol. 
In essence, the universally composable security model captures the full range of potential attacks and adversaries that the 5G-AKA and Handover protocols may face in the context of MVNO. This allows a more comprehensive and realistic framework for analyzing cryptographic protocols compared to the more abstract and idealized Dolev-Yao model. The detailed comparison between the UC and the Dolev-Yao model has been included in \textbf{Appendix B}.

\section{Security and Privacy Requirements for MVNO}
\label{subsec: Design Goals}

In the context of MVNOs, AKA and HO require specific security and privacy requirements due to the unique challenges and privacy concerns inherent in the MVNO setting. The proposed scheme aims to meet the following security and privacy requirements: 

   \vspace{0.1cm}
    \noindent$\bullet$ \textbf{Mutual Authentication Under MVNO settings (MA+).} 
    
    In general, mutual authentication (MA) ensures the legitimacy of both the network components and the user equipment (UE), which is crucial. This property protects against attacks such as man-in-the-middle (MitM) and impersonation attacks by verifying that all communication parties are genuine and authorised. Now, MA+ mandates that network entities, such as gNB and CN, authenticate the UE and vice versa before rendering any services. The key distinction between the two lies in the user's registration with a different network provider (MVNO), while the network entities providing services are provided by the MNO. Consequently, network entities must ensure users' authenticity without accessing their information, necessitating a method to ensure the authenticity and genuineness of other participants ($\UE$) with whom they communicate. Our scheme achieves  MA+ security to provide a privacy-preserving secure communication environment for all network participants. 
     We provide details of this security experiment in Section \ref{sec: Security Analysis}. We define the authentication goal as follows:
     
     \vspace{0.1cm}
     \noindent\ul{\emph{Establishing privacy-preserving mutual authentication between $\UE$ and $\gNB$ under MA+:}} UE can perform the initial authentication of the broadcast message from the gNB. In this way, we can deal with the fake base station attacks. Besides,  at the end of the execution of the proposed scheme, both the UE (registered with MVNO) and gNB (operated by MNO) are expected to establish mutual trust between themselves.
     \vspace{0.1cm}
     
     \noindent\ul{\emph{Establishing privacy-preserving mutual authentication between $\UE$ and $\CN$ under MA+:}} Even though the UE registers into MVNO, not in CN. However,  at the end of the execution of the proposed scheme, both the UE and CN are expected to establish mutual trust between themselves. In this regard, UE does not need to reveal his/her identity. In this context, a weak agreement means that a participant in the protocol has undergone the protocol with its counterpart. Yet, there is no obligation for them to reach a consensus on any data exchanged or secrets established during the session.

     \vspace{0.1cm}

    \noindent$\bullet$ \textbf{Comprehnsive Privacy (CP)}.

    A privacy-preserving approach needs to go beyond mere anonymity, extending to a model where even the core network remains unaware of the user's identity while retaining the ability to validate their legitimacy. Specifically, this holds true even for roaming users undergoing the handover protocol. 
 
    We define the comprehensive privacy goals as follows:
    
    \vspace{0.1cm}
    
    \noindent\underline{\emph{User Anonymity (UA)}}: Given a UE considers an interaction with AKA/HO session involving CN or gNB or both, no active attacker ($\adv_2$) can recover a user's identity ("SUPI must remain secret"). 

    \vspace{0.1cm}
    \noindent\underline{\emph{Unlinkabilty (UL)}}: Given two $\UE$s identified as $\UE_1$ and $\UE_2$, and considering an AKA/HO session involving either $\UE_1$ or $\UE_2$, it is impossible for attackers  to discern which specific $\UE$ ($\UE_1$ or $\UE_2$) it is interacting with.

    \vspace{0.1cm}
    \noindent$\bullet$ \textbf{Privacy-preserving Revocation (PR)}. 

    It is imperative to guarantee that only authorized users access network resources. However, achieving a balance between user revocation and privacy is widely recognized as a challenge in existing literature \cite{alnashwan_privacy-aware_2023}. In the context of MVNO, it is critical to ensure user anonymity and confidentiality while empowering the network to revoke specific users when necessary.

    \noindent$\bullet$ \textbf{Universal Composability (UC).} 
    
    Universally composable security equips MVNOs to counteract security threats effectively in dynamic and interconnected mobile network environments. The proposed scheme is expected to provide a strong security guarantee by ensuring that the proposed scheme remains secure even when composed of arbitrary other protocols in a more extensive system.

\section{Our Proposed Scheme }
\label{Proposed scheme}

In this section, we introduce our proposed scheme, a structured scheme that unfolds in three phases: Registration, Initial Authentication, and Handover. During the registration phase, there are two sub-categories, MNO registration and MVNO registration, based on the entity responsible for the registration process. When a User Equipment ($\UE$) seeks to join the network, the MVNO facilitates the secure transmission of essential information, including IDs and keys, to the registering participant through a secure channel. However, in the case of gNB and CN, it is the responsibility of the MNO to provide them with the keys and certificates. In the Initial Authentication phase, all users can be authenticated anonymously to the network using Zero-Knowledge Proofs (ZKP), as illustrated in Figure \ref{fig:init-auth}. Simultaneously, the authentication of all gNBs to users is achieved through SanSig certificates generated by the MNO during the registration phase. The continuity of secure services for roaming users is ensured through a user re-authentication through the execution of the Handover phase, as depicted in Figure \ref{fig:Handover}. This iterative handover protocol allows roaming users to seamlessly and securely receive services from the network.

\subsection{System Registration}

In the initial phase of the proposed scheme, crucial parameters are defined, involving entity registration and key generation. User registration is entrusted to the MVNO, while the responsibility for gNB and CN registration rests with the MNO.  In a collaborative effort facilitated by the partnership between the MVNO and MNO, essential information is exchanged between the two entities. As a result, this phase can be subdivided into three main components based on the assigned responsible party and the sharing process between them. This segmentation establishes a clear delineation of roles within the collaborative framework, enhancing the efficiency and clarity of the proposed system.

\noindent\textbf{(1) $\boldsymbol{MNO}$ Setup}: 
The MNO takes on the responsibility of generating key pairs for both the CN and gNBs, ensuring their secure distribution. When the CN sends a setup request to the MNO, the latter generates all the necessary keys for the CN. This encompasses pairs of public/secret keys for digital signature ($s\sk_{\CN},s\pk_{\CN}$) and $\SanSig$ ($\pk_{sig}^{\CN}, \sk_{sig}^{\CN}$) algorithms. The latter is responsible for signing certificates for gNBs within the network. 

Similarly, when a $\gNB$ in the network seeks registration with the MNO, it initiates the process by sending a registration request to the MNO, enclosing pertinent information. Upon receipt of the registration request, the MNO undertakes the authentication of the $\gNB$, and subsequently generates an identity for this base station ($\gid$), along with the associated pairs of keys ($s\sk_{\gNB},s\pk_{\gNB}$), ($\SK_{san}^\gNB, \PK_{san}^\gNB$), and a certificate ($\cert_{G}, \sigma_{G}$) using the $\SanSig$ algorithm ($\sigma_{G} \getsr \SanSig.\Sign(\cert_{G}, \sk_{sig}^{CN}, \pk_{san}^{\gNB}, \adm(\cert_{mod}^{G}) ))$. This certificate encompasses $\gid$, the location information of $\gNB$, and a certificate expiration period ($EXP$), denoting the certificate's validity duration. This comprehensive process ensures secure and authenticated registration of $\gNB$ within the network.

\noindent\textbf{(2) $\boldsymbol {MVNO}$ Setup}:
At the beginning, the MVNO creates a common reference string ($\crs$), its trapdoor $\mathsf{td}$ and commitment key ($\mathsf{ck}$) for the Zero-Knowledge Proof (ZKP) algorithm. This proactive step ensures that all registered users have access to essential information. After that, the MVNO generates all key pairs and identities for users and securely distributes these credentials. 
To use the network services provided by the MVNO, each user must first register on the network. The registration process begins when a User Equipment ($\UE$) selects the MVNO plan and provides necessary information to the MVNO. Upon receipt of the registration request, the MVNO creates a pseudo-identity ($\pid$) for the user, which aligns with the specified ZKP range.
Next, the MVNO hashes the user identity ($H_{\pid}$) and stores it in a designated list ($\mathsf{List}$). Finally, the MVNO transmits the $\pid$, $H_{\pid}$, along with the common reference string ($\crs$), and commitment key ($\ck$) to the registered user, thereby completing this secure registration process.

\noindent\textbf{(3) $\boldsymbol {MVNO-MNO}$ information exchange}:
Due to the collaboration between the MVNO and MNO, crucial information is exchanged to ensure the successful and seamless execution of our proposed scheme. On the MVNO side, this involves sharing the pre-generated $crs$ and the $\mathsf{List}$ (i.e. a list of hashed identities of the registered users), which may undergo updates throughout their partnership. Simultaneously, the MNO contributes by sharing essential information with the MVNO, including the public keys of $\CN$ and $\gNB$s. This exchange ensures that users can verify the authenticity of the network, establishing a robust foundation for secure interactions within the system.

\subsection{Initial authentication (MVNO-AKA)}
Each registered user who wants to join the network must execute the initial authentication phase, as shown in Figure~\ref{fig:init-auth}. During the execution of this protocol, the CN generates credentials for new users, which will be used in the subsequent Handover protocols:

\begin{figure}[htp!]
	\centering
	\begin{adjustbox}{max width=0.45\textwidth, max height=\textheight}
		\fbox{
	\begin{tikzpicture}[yscale=-0.34,>=latex]
    \tikzstyle{every node}=[font=\large]
	\edef\InitX{0}
	\edef\ArrowLeft{1}
	\edef\ArrowCenter{6}
	\edef\ArrowRight{11}
	\edef\RespX{12}
	\edef\Y{0}

\node [rectangle,draw=black, inner sep=5pt, align=left] at (\InitX+2.7,\Y) {$\ck,\crs$};

\node [rectangle,draw=black, inner sep=5pt, align=left] at (\InitX+4.1,\Y) {$H_{\pid_i}$};

\node [rectangle,draw=black, inner sep=5pt, align=left] at (\InitX+5.7,\Y) {$H_{\pid_{i+1}}, ...$};

\node [rectangle,draw=black, inner sep=5pt, align=left] at (\InitX+7.30,\Y) {$H_{\pid_n}$};

\NextLine
 \Separatorb{}
  \NextLine[3]
\node (rect) at (1,5.8) [label={[anchor=north]:\textcolor{red}{MVNO's User}},draw,thick,minimum width=5cm,minimum height=2.4cm, color=red, dashed] {};

 \node [right]at (\InitX,\Y) {\textbf{UE}};
	\node [rectangle,draw=black, fill=gray!50,inner sep=5pt, align=left] at (\InitX+1,\Y+3) {$ \ck,\crs, \pid, \pk_{\CN}$\\ $, \pk_{san}^{g}, \pk_{sig}^{\CN},s\sk_u,s\pk_{\CN}$};

  \node [right]at (5,\Y) {\textbf{gNB}};
	\node [rectangle,draw=black, fill=gray!50,inner sep=5pt,align=left] at ((5.5,\Y+3) {$\sk_{san}^{g}, \pk_{san}^{g},s\sk_g,$\\$\cert_G,\sigma_G, \pk_{sig}^{\CN}$};
	
  \node [left]at (\RespX,\Y) {\textbf{CN}};
 \node [rectangle,draw=black, fill=gray!50,inner sep=5pt,align=left] at (\RespX-1,\Y+3) {$s\sk_{\CN},\crs,\ck$};

  \node (rect) at (8.1, 5.8) [label={[anchor=north]:\textcolor{red}{Operated by MNO}},draw, thick, minimum width=8.6cm, minimum height=2.4cm, color=red, dashed] {};

    \NextLine[6.5]

    \AdversaryAction{$\cert^{G}_{mod} \gets \gid\|\tau$}
    \NextLine[1.4]
	\AdversaryAction{$(\cert_{G}^{*},\sigma^{*}_{G}) \gets \SanSig.\Sanit (\cert_{G},\sigma_{G},\cert^{G}_{mod}, \pk_{sig}^{\CN},\sk_{san}^{g} )$}
	\NextLine[3]
	
	\AdversaryToClient{\framebox[1.1\width]{$\boldsymbol{M_1}:[\cert_{G}^{*}$, $\sigma^{*}_{G}$]}}{}
 \NextLine[1.3]
        \ClientAction{\textbf{check} $\gid \& \tau$}
	\NextLine[1.3]
	\ClientAction{\textbf{abort if} $1 \neq \SanSig.\Verify(\cert_{G}^{*}, \sigma^{*}_{G}, \pk_{sig}^{\CN}, \pk_{san}^{g} )$}
	\NextLine[1.4]
    \ClientAction{$(\PK_u,\SK_u) \gets \KeyGen(\lambda)$}
    \NextLine[1.4]
	\ClientAction{$r \getsr \bits{n}$}
	\NextLine[1.3]
     \ClientAction{$c= \com_{\ck}(\pid,r)$}
	\NextLine[1.3]
     \ClientAction{$\pi_{\ZK} \gets \ZK.\Prove(c,\crs,\pid,\mathsf{List})$}
	\NextLine[1.3]
	\ClientAction{$\sigma \gets\Sign(s\sk_u,\pi_{ZK}\|c\|\PK_u\|\tau_2)$}
	\NextLine[3]
	
	\ClientToAdversary{\framebox[1.1\width]{$\boldsymbol{M_2}:[\pi_{\ZK},c, \PK_u, \tau_2,\sigma]$}}{}

	\NextLine[1.3]
    \AdversaryAction{\textbf{check} $\tau_2$}
    \NextLine[1.3]
	\AdversaryAction{\textbf{abort if} $1 \neq \Verify(s\pk_u,\sigma,\pi_{ZK}\|c \| \PK_u \|\tau_2)$}
     \NextLine[1.3]

    \AdversaryAction{$\sigma^* \gets\Sign(s\sk_g,\pi_{\ZK}\|c\|\PK_u\|\tau_3)$}
		\NextLine[3]	
	\AdversaryToServer{\framebox[1.1\width]{$\boldsymbol{M_3}:[\pi_{\ZK},c,\PK_u,\sigma^*, \tau_3 ]$}}{}
 
     \NextLine[1.3]
    \ServerAction{\textbf{check} $\tau_3$}
    \NextLine[1.3]
	\ServerAction{\textbf{abort if} $1 \neq \Verify(s\pk_g,\sigma^*,\pi_{\ZK}\|c \| \PK_u \|\tau_3)$}
 \NextLine[1.3]
	\ServerAction{\textbf{abort if} $1 \neq \ZK.\Verify(\crs,\pi_{\ZK},c )$}
  \NextLine[1.3]
	\ServerAction{$\UID_i\getsr \mathbb{P}$}
	\NextLine[1.3]
        \ServerAction{$H_{\UID_i}\gets Hash(\UID_i)$}
        \NextLine[1.3]
    \ServerAction{$\sigma_U \gets\Sign(s\sk_{\CN}, \UID\|\tau_4)$}

	\NextLine[3]
 
	\ServerToClient{\framebox[1.1\width]{$\boldsymbol{M_{4}}$:$\Enc_{\PK_u}\{\sigma_{U}\|\UID\|\tau_4\}$}}{$via \:\gNB$}

	\NextLine[1.6]
    \ClientAction{$ \sigma_{U}\|\UID\|\tau_4\gets \Dec_{\SK_u}\{M_4\}$}
    \NextLine[1.3]
    \ClientAction{\textbf{check} $\tau_4$}
    \NextLine[1.3]
	\ClientAction{\textbf{abort if} $\Verify(\UID\| \tau_4,\sigma_{U},s\pk_
 {\CN})\neq1$}
	\NextLine[1.3]

	\end{tikzpicture}
 
}
	\end{adjustbox}
	\caption{UC-based Initial Authentication protocol for MVNO}
	\label{fig:init-auth}
\end{figure}
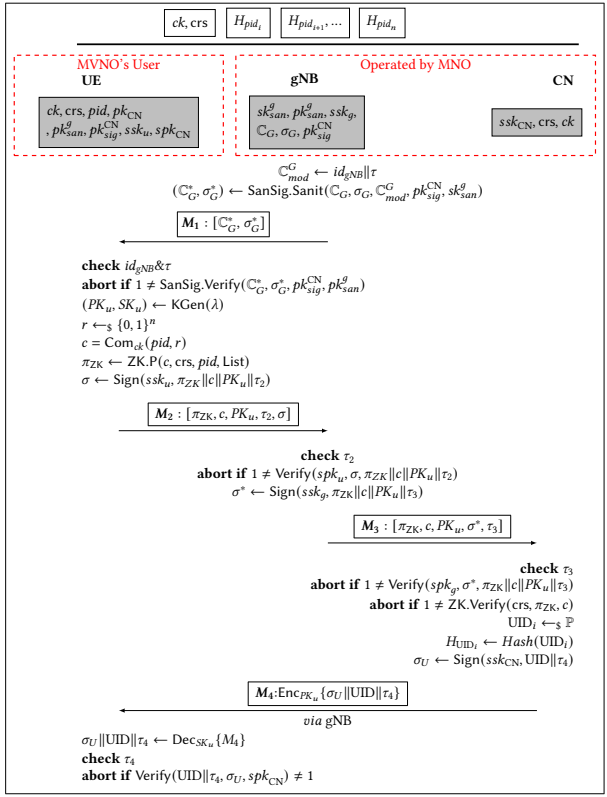

\begin{itemize}[leftmargin=*]
\item[] \textbf{Step 1}:  $\gNB \rightarrow \UE$. $\mathbf{M_{1}}$:[$\cert_{G}^{*},\sigma^{*}_{G}$]:- In the first step, $\gNB$ utilizes the $\SanSig.\Sanit(.)$ algorithm to sanitize their certificate to include their identity ($\gid$) and a timestamp to ensure message integrity and to prevent the known MITM and DOS attacks. The updated/sanitized certificate with its signature ($\sigma$) is sent to the $\UE$ via $M_1$.

\item[] \textbf{Step 2}:  $\UE \rightarrow \gNB$. $\mathbf{M_{2}}$:$[\pi_{ZK},c, \PK_u,\tau_2,\sigma]$:- Upon receiving $M_1$, $\UE$ first checks the timestamp and $\gid$, and verifies the signature using $(\SanSig. \Verify())$. If all verifications hold, then $\UE$ computes a pair of asymmetric keys $(\PK_u,\SK_u)$ for encryption. Next, the $\UE$ samples randomly ($r$), computes a commitment of their $H_{\pid}$ and a zero-knowledge proof ($\pi_{\ZK}$) for the $\LANG_{\mathsf{zk}} = \{ (c, \mathsf{List}) \ | \ \exists \wit:= (\pid, r) \text{ s.t } c= \mathsf{Com}_{\mathsf{ck}}(\pid,r) \land H_\pid \in \mathsf{List}\}$ where $\mathsf{List}$ is list of the all hashes of the user identity, $\mathsf{List} = \{H_{\pid_1}, \cdots , H_{\pid_n} \}$. Finally, the user signs all the previous computations and sends them to  $\gNB$.

\item[] \textbf{Step 3}:- $\gNB \rightarrow CN$. $\mathbf{M_3}:[\pi_{ZK},c,\PK_u, \tau_3, \sigma^*]$: After receiving the message $M_2$, $\gNB$ first checks the timestamp and verifies the signature $\sigma$ to check the message integrity. If both verifications hold, $\gNB$ signs the content of $M_2$ using his singing key ($s\sk_g$). Then $\gNB$ forwards the generated signature ($\sigma^*$) along with $M_2$ to the $CN$.

\item[] \textbf{Step 4}: $CN \rightarrow \UE$. $\mathbf{M_{4}}: [\Enc_{\PK_u}\{\sigma_{U}\|\UID\|\tau_4\}] $ :-
Upon receiving $M_3$, the $CN$ verifies both ($\sigma^*$) and ($\pi_{ZK}$) using signature and ZKP verification algorithms, respectively. If the verification holds, the $CN$ computes a universal user ID $(\UID_i)$, which will be the user's identifier during HOs. Then CN hash and signs $(\UID_i)$ along with a timestamp ($\tau$) using $\Sign(.)$ algorithm. CN then stores $\UID_i$ and encrypts user certificate $\UID_i$, $\sigma_U$ and timestamp using ($\PK_u$). Finally, the $CN$ sends $M_{4}$ to the $\UE$ via $\gNB$. Then, the user is responsible for verifying $M_{4}$ and the signature.

\end{itemize}

\begin{remark}
    If a symmetric session key is needed between the UE and gNB for future communication, the CN can distribute one to both entities within message $M_4$. This message will encapsulate the encryption of the identical session key twice: once encrypted with the gNB's public key and the other with the UE's public key.
\end{remark}

\subsubsection{Integration of the Proposed Initial Authentication Protocol with 5G AKA} \label{sec: 5gaka}

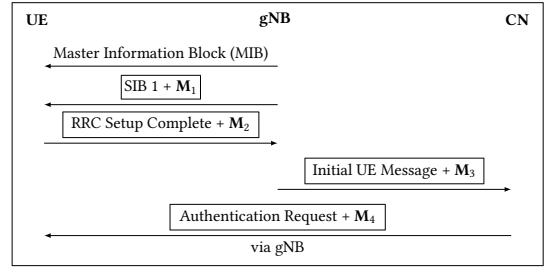
\begin{figure}[htp!]
	\centering
    \vspace{-0.3cm}
	\begin{adjustbox}{max width=0.40\textwidth, max height=\textheight}
		\fbox{
	\begin{tikzpicture}[yscale=-0.33,>=latex]
    \tikzstyle{every node}=[font=\large]
	\edef\InitX{0}
	\edef\ArrowLeft{1}
	\edef\ArrowCenter{6}
	\edef\ArrowRight{11}
	\edef\RespX{12}
	\edef\Y{0}





\NextLine
  \NextLine[3]
 \node [right]at (0.5,\Y) {\textbf{UE}};

  \node [right]at (5.5,\Y) {\textbf{gNB}};
	
  \node [left]at (\RespX-0.5,\Y) {\textbf{CN}};


    \NextLine[2.5]
    \AdversaryToClient{Master Information Block (MIB)}{}
    \NextLine[1.5]
    \AdversaryToClient{\framebox[1.1\width]{SIB 1 + $\textbf{M}_1$}}{}
    \NextLine[1.5]
    \ClientToAdversary{\framebox[1.1\width]{RRC Setup Complete + $\textbf{M}_2$}}{}
    \NextLine[2]
    \AdversaryToServer{\framebox[1.1\width]{Initial UE Message + $\textbf{M}_3$}}{}

    \NextLine[2]
    \ServerToClient{\framebox[1.1\width]{Authentication Request + $\textbf{M}_4$}}{via gNB}

    \NextLine[1]

	\end{tikzpicture}
 
}
	\end{adjustbox}
	\vspace{-0.3cm}
    \caption{Integration of our Initial Authentication Protocol with 5G-AKA}
	\label{fig:5G-AKA}
\end{figure}

This section shows how our proposed initial authentication scheme can be integrated with the conventional 5G-AKA. As shown in Figure \ref{fig:5G-AKA}, our $M_1$ can be integrated into SystemInformationBlockType1 (SIB1) in the current 5G protocol. By authenticating broadcast messages, the UE can verify the authenticity of the gNB before connecting to it. Since an attacker cannot forge the signature, this method helps prevent fake base station attacks. Then, $M_2$ is integrated into the RRC Setup Complete message. We extend the Initial UE Message, an NGAP (Next Generation Application Protocol) message between gNB and CN, to transport $M_3$. The response from CN, $M_4$, is sent to UE as a response to $M_3$ inside of the authentication request message. Since our proposed scheme only extends the existing 5G control-plane messages, it is fully compatible with the current 5G authentication scheme. If the UE or the network does not support the new scheme, we can downgrade to the existing 5G-AKA. In our proposed scheme, the UE is not required to send its identity in plaintext over-the-air, and an external attacker cannot forge plaintext 5G control-plane messages to acquire sensitive UE information or perform DoS attacks. Note that this integration does not introduce any new 5G protocol messages and reuses only the extension of existing messages. Hence, our proposed solution will not affect the legacy UEs incapable of supporting our solutions. 

\subsection{UC Secure MVNO Handover Scheme (MVNO-HO)}
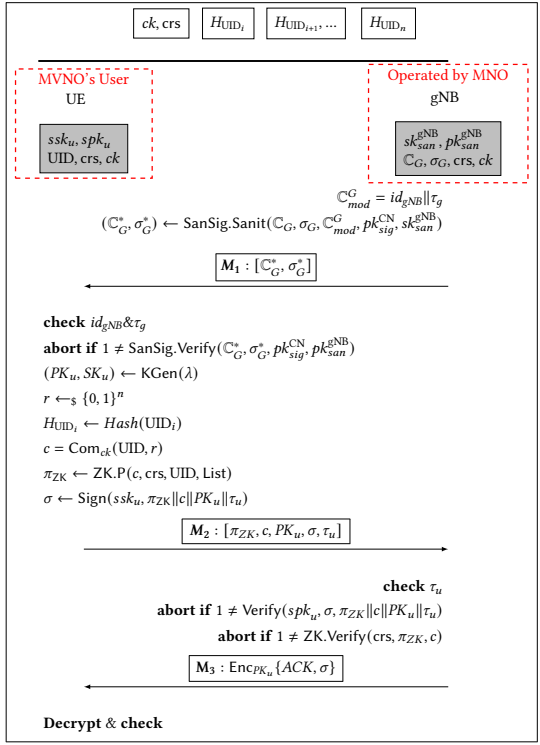
\begin{figure}[htbp]
	\centering
    \vspace{-0.3cm}
	\begin{adjustbox}{max width=0.4\textwidth, max height=\textheight}
		\fbox{
	\begin{tikzpicture}[yscale=-0.55,>=latex]
    \tikzstyle{every node}=[font=\large]
	\edef\InitX{0}
	\edef\ArrowLeft{1}
	\edef\ArrowCenter{6}
	\edef\ArrowRight{9}
	\edef\RespX{9}
	\edef\Y{0}
	

\node [rectangle,draw=black, inner sep=5pt, align=left] at (\InitX+2.7,\Y) {$\ck,\crs$};

\node [rectangle,draw=black, inner sep=5pt, align=left] at (\InitX+4.17,\Y) {$H_{\UID_i}$};

\node [rectangle,draw=black, inner sep=5pt, align=left] at (\InitX+5.88,\Y) {$H_{\UID_{i+1}}, ...$};

\node [rectangle,draw=black, inner sep=5pt, align=left] at (\InitX+7.69,\Y) {$H_{\UID_n}$};
\NextLine
 \Separatorb{}
 \NextLine[2]
 
    \node (rect) at (1,4) [label={[anchor=north]:\textcolor{red}{MVNO's User}},draw,thick,minimum width=3cm,minimum height=2.4cm, color=red, dashed] {};
    \node [right]at (\InitX+0.5,\Y) {\textbf{$\UE$}};
    \node [rectangle,draw=black, fill=gray!50,inner sep=5pt, align=left] at (\InitX+1,\Y+2) {$s\sk_u, s\pk_u$\\$\UID,\crs,\ck$};
    
    \node (rect) at (9, 4) [label={[anchor=north]:\textcolor{red}{Operated by MNO}}, draw,thick, minimum width=3.5cm, minimum height=2.6cm, color=red, dashed] {};
    \node [right]at (\RespX-0.5,\Y) {\textbf{$\gNB$}};
    \node [rectangle,draw=black, fill=gray!50,inner sep=5pt,align=left] at ((\RespX,\Y+2) {$\sk_{san}^{\gNB}, \pk_{san}^{\gNB}$\\$\cert_G,\sigma_G,\crs,\ck$};

	\NextLine[4]
 
    \ServerAction{$\cert^{G}_{mod}= \gid\|\tau_g$}
    \NextLine
	\ServerAction{$(\cert_{G}^{*},\sigma^{*}_{G}) \gets \SanSig.\Sanit (\cert_{G},\sigma_{G},\cert^{G}_{mod}, \pk_{sig}^{\CN},\sk_{san}^{\gNB} )$}
	\NextLine[2]
	
	\ServerToClient{\framebox[1.1\width]{$\boldsymbol{M_1}:[\cert_{G}^{*}$, $\sigma^{*}_{G}$]}}{}
 
	\NextLine

    \ClientAction{\textbf{check} $\gid \& \tau_g$}
    \NextLine
	\ClientAction{\textbf{abort if} $1 \neq \SanSig.\Verify(\cert_{G}^{*}, \sigma^{*}_{G}, \pk_{sig}^{\CN}, \pk_{san}^{\gNB} )$}
	\NextLine
    \ClientAction{$(\PK_u,\SK_u) \gets \KeyGen(\lambda)$}
	\NextLine
	\ClientAction{$r \getsr \bits{n}$}
	\NextLine
     \ClientAction{$H_{\UID_i}\gets Hash(\UID_i)$}
        \NextLine
     \ClientAction{$c= \com_{\ck}(\UID,r)$}
	\NextLine
     \ClientAction{$\pi_{\ZK} \gets \ZK.\Prove(c, \crs,\UID,\mathsf{List})$}
	\NextLine
	\ClientAction{$\sigma \gets\Sign(s\sk_u,\pi_{\ZK}\|c\|\PK_u\|\tau_u)$}
 
	\NextLine[1.5]
	\ClientToServer{\framebox[1.1\width]{$\boldsymbol{M_2}:[\pi_{ZK},c, \PK_u,\sigma, \tau_u]$}}{}
	\NextLine
    \ServerAction{\textbf{check} $\tau_u$}
	\NextLine
    \ServerAction{\textbf{abort if} $1 \neq \Verify(s\pk_u,\sigma,\pi_{ZK}\|c \| \PK_u\| \tau_u)$}
    \NextLine
    \ServerAction{\textbf{abort if} $1 \neq \ZK.\Verify(\crs,\pi_{ZK},c )$}
	\NextLine[1.5]
 \ServerToClient{\framebox[1.1\width]{$\mathbf{M_{3}}:\Enc_{\PK_u}\{ACK, \sigma\}$}}{}
	\NextLine
  \ClientAction{\textbf{Decrypt $\&$ check}}
	\NextLine[1.5]
	
	\end{tikzpicture}
}
	\end{adjustbox}
    \vspace{-0.3cm}
	\caption{Proposed UC Secure MVNO Handover protocol}
	\label{fig:Handover}
    \vspace{-0.3cm}
\end{figure}

Users who complete initial authentication and want to roam between small cells must execute this protocol. As part of this protocol, users will utilize the Zero-Knowledge Proof (ZKP) and the unique identities ($\UID$) generated from the previous protocol to verify their authenticity to the $\gNB$s. Subsequently, the $\gNB$s will cross-check these identities with the list of identities created by the $CN$ during the previous protocol. The Handover protocol is described below and illustrated in Figure \ref{fig:Handover}.

\begin{itemize}[leftmargin=*]
\item[]\textbf{Step 1:}  $\gNB \rightarrow \UE$. $\mathbf{M_{1}}$:[$\cert_{G}^{*},\sigma^{*}_{G}$]:- This step is identical to the first step of the initial authentication protocol. 
    
\item[]\textbf{Step 2:} $\UE \rightarrow \gNB$. $\boldsymbol{M_2}:[\pi_{ZK},c, \PK_u,\sigma, \tau_u]$ :-
    Upon receiving $M_1$, $\UE$ first checks the timestamp and $\gid$, and verifies the signature using $(\SanSig. \Verify())$. Assuming both verifications are successful, then $\UE$ computes a pair of asymmetric keys $(\PK_u,\SK_u)$ for encryption. Next, the $\UE$ samples randomly ($r$), computes a commitment of their $H_{\UID}$ and a zero-knowledge proof ($\pi_{ZK}$) for the $\LANG_{\mathsf{zk}} = \{ (c, \mathsf{List}) \ | \ \exists \wit:= (\UID, r) \text{ s.t } c= \mathsf{Com}_{\mathsf{ck}}(\UID,r) \land H_\UID \in \mathsf{List}\}$ where $\mathsf{List} = \{H_{\UID_1}, \cdots , H_{\UID_n} \}$. Finally, the user signs all the previous computations and sends them to  $\gNB$.

\item [] \textbf{Step 3}: $\gNB \rightarrow UE$ $\mathbf{M_{3}}$:$\Enc_{\PK_u}\{ACK, \sigma\}$ :-
    Upon receipt of $M_{2}$, $\gNB$ checks the timestamp and verifies the signature and identity of $\UE$ using signature and ZKP verification algorithms, respectively. If both verifications hold, $\gNB$ encrypts $ACK$ and $\sigma$ using the user's public key and sends it via $M_{3}$ to UE. Finally, after receiving $M_{3}$, the user decrypts the message and checks the integrity of $\sigma$. Details of this protocol are depicted in Figure \ref{fig:Handover}.
\end{itemize}

\subsubsection{Integration of the Proposed Scheme with 5G-Handover}

\begin{figure}[htp!]
	\centering
	\begin{adjustbox}{max width=0.4\textwidth, max height=\textheight}
		\fbox{
    \begin{tikzpicture}[yscale=-0.33,>=latex]
    \tikzstyle{every node}=[font=\large]
    \edef\InitX{0}
    \edef\ArrowLeft{1}
    \edef\ArrowCenter{6}
    \edef\ArrowRight{11}
    \edef\RespX{12}
    \edef\Y{0}
    
    \NextLine
    
    \NextLine[3]
    
    \node [right]at (0.5,\Y) {\textbf{UE}};

    \node [left]at (\RespX-0.5,\Y) {\textbf{gNB}};
    
    \NextLine[2.5]
    \ServerToClient{Master Information Block (MIB)}{}
    \NextLine[1.8]
    
    \ServerToClient{\framebox[1.1\width]{System Information Block Type 1 + \textbf{M\_1}}}{}
    \NextLine[1.8]
    \ClientToServer{\framebox[1.1\width]{RRC Reestablishment Complete + \textbf{M\_2}}}{}
    
    \NextLine[1.8]
    
    \ServerToClient{\framebox[1.1\width]{DL Information Transfer + \textbf{M\_3}}}{}

    \end{tikzpicture}
 
}
	\end{adjustbox}
    \vspace{-0.3cm}
	\caption{Integration of our MVNO Handover Protocol with 5G Handover Scheme}
	\label{fig:5G-HO}
        \vspace{-5mm}
\end{figure}
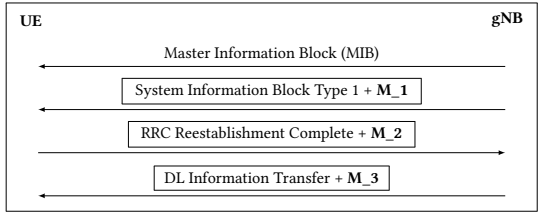

As shown in Figure \ref{fig:5G-HO}, our proposed UC secure MVNO handover protocol can also be integrated with 5G Handover between gNBs. Similarly to Section \ref{sec: 5gaka}, $M_1$ is integrated in the SIB1 message. Thus, the UE can verify the identity of gNB before making the handover decision. $M_2$ is added to the RRC Reestablishment Complete message. Since we don't need communication with CN in this scenario, the gNB directly replies $M_2$ with $M_3$ in an extended DL Information Transfer message. After receiving $M_3$ from gNB, the UE and gNB are mutually authenticated. Our protocol can prevent fake base station attacks on the UE and prevent fake UE attacks from trying to exhaust the resources of the gNB. This design is consistent with our proposed scheme for 5G-AKA, so it is also compatible with the current 5G specifications. 

\subsection{Privacy-preserving Revocation}
\label{Efficient Revocation}
The complex structure of MNO-MVNO networks, the critical need for robust data protection, and escalating security threats necessitate a new approach to user revocation. Our solution addresses this challenge by providing comprehensive privacy during revocation, tackling the unique complexities of multi-party MVNO environments. This approach not only enhances user trust and safeguards personal information but also strengthens overall security without compromising anonymity. It enables the removal of compromised, unauthorized, or non-paying users during authentication and handover processes, significantly improving network efficiency by preventing unnecessary resource consumption. However, implementing such a mechanism while ensuring complete user privacy presents a significant challenge.
Our scheme overcomes this challenge through an innovative approach implemented during the registration phase. By utilizing the shared $\mathsf{List}$ between MNO and MVNO. The procedure involves the MVNO identifying the unique pseudo-ID ($\pid$) associated with the revoked user's identity, hashing it to generate $H_{\pid}$, and then transmitting this $H_{\pid}$ to the core network (CN) for removal from the public list of authorized UEs ($\mathsf{List}$). Subsequently, the CN retrieves the associated Hashed universal ID ($H_{\UID}$) with $H_{\pid}$ to remove it from the handover public list.

\subsection{Instantiating the Primitives}

In this section, we list all primitives used in each part of the proposed construction and explain how to instantiate each primitive using different hardness assumptions.

\noindent \textbf{Non-interactive commitments} are constructed using the following primitives:
\begin{itemize}[leftmargin=0.3cm]
    \item The Pedersen commitment \cite{pedersen1991non} that perfectly hides and computationally binds based on the discrete logarithm (DL) assumption. The UC-secure commitment \cite{abdolmaleki2019framework,abdolmaleki2019dl} from DL Assumption.
    \item The string commitment scheme by Kawachi \textit{et al.}’s \cite{kawachi2008concurrently} based on the SIS assumption \cite{ajtai1996generating}.
    \item The commitment scheme by \cite{jain2012commitments} where the hiding property is based on the Learning Parity with Noise (LPN) assumption, a special case of the Learning With Errors (LWE) assumption \cite{regev2009lattices}.
    \item The commitment scheme \cite{xie2013zero} that is based on Ring-LWE \cite{lyubashevsky2013ideal} instead of LPN, and they build $\Sigma$-protocols from it. Further $\Sigma$-protocols based on (Ring-)LWE encryption schemes were presented by Benhamouda \textit{et al.} \cite{benhamouda2014better}.
    \item The commitment scheme by Baum \textit{et al.} \cite{baum2018more} that relies on the Module-LWE and Module-SIS assumption.
\end{itemize}

\noindent \textbf{Practical non-interactive zero-knowledge protocols} are constructed using the following primitives:
\begin{itemize} [leftmargin=0.3cm]
    \item Depending on the computation cost of the prover, verifier, and the communication complexity, there are several practical transparent zero-knowledge argument schemes based on the discrete logarithm assumption. Spartan \cite{setty2020quarks} with compact proof proof size (tens of KBs) and based on bilinear pairing. The Bulletproofs \cite{bunz2018bulletproofs} and Supersonic \cite{bunz2020transparent} are based on discrete-log and group of unknown order with the smallest proof size (1-2 KBs).
    \item The are several practical post-quantum transparent zero-knowledge argument schemes such as Orien \cite{xie2022orion}, Brakedown \cite{golovnev2021brakedown}, Aurora \cite{ben2019aurora}, and Ligero \cite{ames2017ligero}. Orion has the fastest prover among all schemes. This is slightly faster than Brakedown and is 20 times faster than Ligero and 142 times faster than Aurora because of the linear prover time.
    The proof size of Orion is significantly smaller than Brakedown and Ligero. But Aurora has the most compact proof size.
\end{itemize}

\noindent \textbf{Practical sanitizable signatures} are constructed using the following primitives:
Sanitizable signatures offer diverse construction methods, such as employing chameleon hashes and a standard digital signature scheme, as demonstrated in \cite{ateniese_sanitizable_2005}. Another approach involves using two types of signatures: the conventional digital signature and a group signature, as outlined in \cite{brzuska_unlinkability_2010}. Additionally, the creation of a sanitizable signature can involve utilizing two conventional digital signature schemes, such as RSA-based signature and ECDSA, as depicted in \cite{brzuska_santizable_2009,brzuska_non-interactive_2013, brzuska_efficient_2014}. It's important to note that the performance and security of the SanSig may vary based on the underlying construction and primitives used.

\section{Security Analysis}
\label{sec: Security Analysis}
In this section, we provide the security proof of our construction. We first define $\mathcal{F}_{mvno}$ as the ideal functionality of our proposed scheme below. Note that besides the session identifier $sid$, the functionality now
takes another unique "identifier" $cid$, which may be used if a sender
sends to the same receiver multiple times within a session. We assume that the combination of $sid$ and $cid$ is globally unique.

$\mathcal{F}_{mvno}$ parameterized by a message space $\mathcal{M}$ and interact with adversary $\textbf{Sim}$ and parties $P_1, P_2, ... ,P_n$ as follows:

\vspace{-0.1cm}
\begin{itemize}[leftmargin=0.3cm]
    \item Upon receiving $(\mathsf{message_1} ,sid, cid, P_i, P_j , M_1)$ from $P_i$, it
    proceeds as follows: if a tuple $(sid, cid, \cdots)$ with the same $(sid, cid)$ was previously recorded, do nothing. Otherwise, record $(sid, cid, $ $ P_i, P_j , M_1 )$ and send $(\mathsf{message_1}, sid, cid, P_i, P_j)$ to $P_j$ and $\simulator$.

     \item Upon receiving $(\mathsf{message_k} ,sid, cid, P_i, P_j , M_2)$ from $P_i$ for $k \in \{2,3\}$, it
    proceeds as follows: if a tuple $(sid, cid, \cdots)$ with the same $(sid, cid)$ was previously recorded, record $(sid, cid, P_i, P_j, M_k)$ and send $(\mathsf{receipt}, sid, cid, P_i, P_j)$ to $P_j$ and $\simulator$. Otherwise, do nothing.
    \item Upon receiving $(\mathsf{message_4} ,sid, cid, P_i, P_j , M_4)$ from $P_i$, it
    proceeds as follows: if a tuple $(sid, cid, \cdots)$ with the same $(sid, cid)$ was previously recorded, do nothing. Otherwise, record $(sid, cid, $ $ P_i, P_j , M_4 )$ and send $(\mathsf{message_4}, sid, cid, P_i, P_j)$ to $P_j$ and $\simulator$.
    
    \item Upon receiving $(\mathsf{corrupt}, sid, cid)$ from the adversary, send $M$ to the adversary if there is already an entry $(sid, cid, P_i, P_j , M )$. Change the record to $(sid, cid, P_i, P_j , M^*)$, if the adversary provides some $M^*$ and $(\mathsf{receipt}, sid, cid, P_i, P_j)$ has not yet been written on $P_j$ ’s output tape.
\end{itemize}

\vspace{-0.1cm}
\subsection{Formal security analysis}

In this section, we analyse the security of the proposed scheme. In this regard, we consider formal security analysis that primarily revolves around the Universal Composability (UC) framework \cite{canetti2001universally}, which is inherently designed to offer strong, composable security guarantees that extend to the real and ideal models of protocol execution. This framework provides a robust structure for analysing the security properties of cryptographic protocols in a way that is preserved even when these protocols are composed of others in complex systems. While tools like ProVerif \cite{problanchet2016modeling} and Tamarin \cite{tenmeier2013tamarin} are indeed powerful for automating the verification of security properties, particularly in cryptographic protocols, their use is primarily tailored towards protocols that need to be verified against specific properties such as secrecy and authentication in symbolic models. Our manual analysis allows for detailed, nuanced handling of the specifics of the UC model, which are often only directly supported by these automated tools with considerable customisation and significant manual efforts.
Thus, while ProVerif \cite{problanchet2016modeling} and Tamarin \cite{tenmeier2013tamarin} are invaluable in many contexts, the security validation in our paper is most appropriately addressed through rigorous manual proofs within the UC framework. This ensures accurate and context-specific security assurance that aligns with our protocol’s innovative aspects.

We consider a sequence of \textit{hybrid} games between the real and ideal worlds. This is a general approach that one can follow to prove the security of a commitment scheme in the UC model. The game starts with the real game, adversary $A$ interacts with real parties, and ends with the ideal game. In the ideal game, we build $SIM$ that interfaces between adversary $A$ and ideal functionality $F_{mvno}$.

\textit{$Game_0$}: This is the original real game that corresponds to the real world in the model. This game executes the real protocol between committer $P_i$ and receiver $P_j$. The environment $\mathcal{Z}$ chooses the input for the honest committer $P_i$, and $\mathcal{Z}$ receives the output of the honest committer. In our framework, there is an adversary $\mathcal{A}$ that aims to attack the real protocol in the real world by corrupting some parties $P$ and listening to all flows from parties. In that case, $\mathcal{A}$ can read the corrupted party's current inner state and fully control it. In our security game, environment $\mathcal{Z}$ can control adversary $\mathcal{A}$ and see all communication messages from all parties and also all of $\mathcal{A}$'s interactions with other parties.

\textit{$Game_1$}: We consider that the adversary $\mathcal{A}$ controls UE. In the setup phase of this game, simulator $\textbf{Sim}$ chooses the $crs$, its trapdoor $\mathsf{td}$ and utilises the $\mathsf{SanSig.Sanit()}$ algorithm to sanitise the certificate to include their identity and a timestamp. After that, simulator $\textbf{Sim}$ sends a certificate with its signature $\sigma_G^*$ to the UE via $M_1$. 
Upon receiving message $M_2$ from UE, simulator $\textbf{Sim}$ first check the $\tau_2$ and verify the Message using zero-knowledge proof. After that, simulator $\textbf{Sim}$ randomly generates a UID$_i$ and computes the hash value $H_{UID_i}$. Finally, simulator $\textbf{Sim}$ encrypt message $M_4$ and send it to the User.

\textbf{Lemma 1}. \textit{If} $\Pi$\textit{ = (KGen, SanSig, Enc, Dec, Sign, Verify)}, \textit{the labelled signature is EUF-CMA secure, the labelled ZKP holds commitment sound and binding, the output of} $\mathcal{Z}$ \textit{in} $Game_0$ \textit{and} $Game_1$ \textit{is computationally indistinguishable.}

\textit{Proof.} In $Game_1$, we consider two possible scenarios and split our proof into two cases. We first observe that SIM reveals verified results after some party $P_i^1$ open commitment to message $M_1$. We assume that \textbf{bad} defines the case that sender $P_i^1$ successfully generates a valid $H_{pid}$, which means it can generate a valid $\pi_{ZK}$. The bad happens with a negligible probability because the commitment generated by the ZKP algorithm has the sound and binding property. In the second scenario, we assume that \textbf{bad} defines the case that sender $P_i^2$ successfully generates a valid $\sigma$. The bad happens with a negligible probability because the signature scheme is EUF-CMA secure.
Hence, from the proof above, the bad cases happen only with a negligible probability, and two games $Game_0$ and $Game_1$ are computationally indistinguishable in a view of $\mathcal{Z}$.

\textit{$Game_2$}: In this game, we consider that the adversary $\mathcal{A}$ controlled gNB and CN, which means the UE doesn't trust them. Upon the simulator $\textbf{Sim}$ receiving message $M_1$ from the gNB, the simulator $\textbf{Sim}$ first check the $id_{gNB}\&\tau$. After that, the simulator $\textbf{Sim}$ verifies the sanitizable signature using SanSig.Verify. After generate the secret key $SK_{Sim}$ and public key $SK_{Sim}$, the simulator $\textbf{Sim}$ generate $r\leftarrow_\$\{0,\ 1\}$. Hence the simulator $\textbf{Sim}$ don't know the $pid$, it generates the commitment $c$ from $c$ = Com$_{ck}(0, r)$. And than the simulator $\textbf{Sim}$ generates the proof using $\pi_{ZK}\leftarrow$ ZP.SIM$(\mathsf{td},c,crs,0)$. Finally, the simulator $\textbf{Sim}$ sends message $M_2$ to the gNB.

\textbf{Lemma 2}. \textit{If} $\Pi$\textit{ = (KGen, ZK.P, Sign, SanSig, Verify)}, \textit{the labelled signature is EUF-CMA secure, the labelled ZKP holds zero-Knowledge property and the commitment has hiding property, the output of} $\mathcal{Z}$ \textit{in} $Game_1$ \textit{and} $Game_2$ \textit{is computationally indistinguishable.} 

\textit{Proof.} In $Game_2$, we can observed that after $P_i$ open commitment to message $M$, simulator SIM reveals verified results $V_{\sigma}$. Suppose that \textbf{bad} defines the case that receiver $P_j$ successfully get value $H_{pid}$. The \textbf{bad} happens with a negligible probability due to the zero-knowledge property and the hiding property of the commitment. Therefore, the proof generated by the simulator $\textbf{Sim}$ is indistinguishable from the real proof. Also, due to the unlinkability and the unforgeability of the signature mechanism, the output of the signature for both simulator $\textbf{Sim}$'s zero-knowledge proof and real proof are indistinguishable.

Hence, from the proof above, the bad case happens only with a negligible probability and two games $Game_1$ and $Game_2$ are computationally indistinguishable in a view of $\mathcal{Z}$.

\textit{$Game_3$}: This game corresponds to the ideal world in the CRS model. In an ideal world, there exists an ideal function \textit{$F_{mvno}$} and an honest task. Parties in the ideal world simply pass inputs from environment Z to the ideal world function and vice versa. In an ideal world, an ideal honest party interaction has only environment Z and ideal functionality. In this game, the ideal world adversary $\textbf{Sim}$ proceeds following functions:

\vspace{-0.2cm}
\begin{itemize}[leftmargin=0.3cm]
    \item \textit{Initialisation step:} $\simulator$ chooses the $\crs$, its $\tc$, and $\mathsf{ck}$.

    \item \textit{Simulating the communication with $\mathcal{Z}$:} Every input value that Sim receives from $\mathcal{Z}$ is written on $\adv$’s input tape (as if coming from $\mathcal{Z}$) and vice versa.

    \item \textit{Simulating the the first round when sender $P_i$ is honest}: Upon receiving the receipt message $(\mathsf{receipt}, sid, cid, P_i, P_j )$ from $\mathsf{F}_{mvno}$, $\simulator$ computes $M_k$ like a honest party and sends ($\mathsf{message_1}, sid, cid$, $M_k)$ to $P_j$.

    \item \textit{Simulating the second round when sender $P_i$ is honest}: Upon receiving the receipt message $(\mathsf{receipt}, sid, cid, P_i, P_j )$ from $\mathsf{F}_{mvno}$, $\simulator$ computes $M_2$ by $\mathsf{Com}_{\mathsf{ck}} (0, r)$ for randomly chosen r, and run the ZK simulator to compute the proof $\pi_{zk}$ (with using the trapdoor $\tc$) and sends ($\mathsf{message_2}, sid, cid, M_2$) to $P_j$.

    \item \textit{Simulating the the round $k \in \{3,4\}$ when sender $P_i$ is honest}: Upon receiving the receipt message $(\mathsf{receipt}, sid, cid, P_i, P_j )$ from $\mathsf{F}_{mvno}$, $\simulator$ computes $M_1$ like a honest party and sends ($\mathsf{message_1}$, $sid, cid, M_1)$ to $P_j$.

    \item   \textit{Simulating adaptive corruption of Pi after the round k}: When $P_i$ is corrupted, $\simulator$ can immediately read ideal $P_i$’s inner state and obtain $M$ . Then, $\simulator$ produces $M_k$ as in the case of the round $k+1$ when $P_i$ is honest and outputs $( sid, cid, M_k )$ to the $P_j$.

    \item \textit{Simulating the commit phase when committer $\hat{P}_i$ is corrupted and the receiver $P_j$ is honest}: After receiving $(\mathsf{message}_k, sid, $ $cid, M_k)$ from $\hat{P}_i$ controlled by $\adv$ in the round k, $\simulator$ runs the extractor of ZK and compute  $M'$, and sends $(\mathsf{message}, sid,cid, $ $ P_j , M')$ to $\mathsf{F}_{mvno}$.

    \item \textit{Simulating adaptive corruption of $P_j$ after the round k but before the verifying phase}: When $P_j$ is corrupted, $\simulator$ simply outputs $(sid, cid, M_k)$.
\end{itemize}

By the construction of the above functions, $Game_3$ is identical to the $Game_2$.

\section{Comparison with Existing Work}
\label{sec: Concurrent Works}

\begin{table}[t]
    \caption{Comparison with the state-of-the-art.}
    \vspace{-0.3cm}
    \centering
    \resizebox{\columnwidth}{!}{%
    \begin{tabular}{|l |c |  c c c c c|} 
        \toprule[1.3pt]

         \multirow{2}{*}{\diagbox[width=4cm]{\textbf{Schemes}}{\textbf{Features}}} & \multirow{2}{*}{\textbf{SMCT}} &  \multirow{2}{*}{\textbf{MA+}} & \multicolumn{2}{|c|}{\textbf{CP}} & \multirow{2}{*}{\textbf{UC}} & \multirow{2}{*}{\textbf{PR}} \\ 
         &  & &  \multicolumn{1}{|c|}{\textbf{AN}} & \multicolumn{1}{|c|}{\textbf{UL}} &  &  \\
         
        \toprule[1.5pt]
         Conventional-5G\cite{3rd_generation_partnership_project_3gpp_security_2020}&5G& \xmark &  \cmark&  \xmark &  \xmark& \xmark \\  
        \hline

          $5G-AKA' $ \cite{wang_privacy-preserving_2021}& 5G & \xmark&  \cmark  &  \cmark &  \xmark& \xmark
          \\\hline

         PGPP \cite{schmitt2021pretty} &  MVNO & \xmark &  \cmark &  \cmark &  \xmark& \xmark
        \\\hline

         AAKA \cite{yu_aaka_2024}&  4G. 5G & \xmark&  \cmark &  \cmark &  \xmark & \xmark
         \\
        \hline

        Thick Model and SM-DP+ &  MVNO*  & \cmark&  \cmark&  \xmark &  \xmark&  \xmark 
        \\
         \hline

         MVNO-AKA (Ours) &  MVNO* & \cmark&  \cmark&  \cmark &  \cmark&  \cmark 
        \\
         \hline
        MVNO-HO (Ours) &  MVNO* & \cmark&  \cmark&  \cmark &  \cmark&  \cmark 
        \\

    \midrule[1.5pt]
    \multicolumn {7}{|c|}{\textbf{SMCT}: Supported Mobile Communication Type; }
    \\
    \multicolumn {7}{|c|}{ \textbf{MA+}: MA within MVNO environment; \textbf{AN}:Anonymity; \textbf{UL}: Unlinkability}
        \\
    \multicolumn {7}{|c|}{ \textbf{CP}: Comprehnsive Privacy, \textbf{UC}: Universal Compositions Security;  }
    \\
    \multicolumn {7}{|c|}{\textbf{PR}: Privacy-preserving Revocation, \textbf{MVNO*}: Applicable for MVNO,}
    \\
    \multicolumn {7}{|c|}{ 4G, 5G and next-generation mobile communication}
    \\
    \multicolumn {7}{|c|}{ \textbf{Thick Model and SM-DP+}: Thick model and an SM-DP+ that rotates IMSIs.}
    \\
    \midrule[1.5pt]
    
    \end{tabular}
    }
    \label{tab:sec.Features}
    \vspace{-5mm}
\end{table}

In this section, we explore state-of-the-art closely aligned with this research and qualitatively compare them with our proposed approach. Considering the privacy aspect, Wang et al. \cite{wang_privacy-preserving_2021} focus on addressing only linkability attacks within the current 5G AKA protocol. However, this protocol does not provide complete privacy protection, such as preventing identity exposures, privacy-preserving mutual authentication, and universal composability within the MVNO environment. Moreover, Wang et al. do not consider secure revocation mechanism which is crucial for the privacy of MVNO users.
Additionally, Schmitt and Raghavan \cite{schmitt2021pretty} have proposed a refactor-based approach, named PGPP, to safeguard user identity and location privacy. The authors introduce a logical entity, termed the PGPP Gateway (PGPP-GW), situated interstitially between the User Plane Function (UPF) and the public Internet. This configuration serves to decouple authentication from connectivity credentials, thereby providing a mechanism for authentication while concurrently preserving user privacy. However, while PGPP-GW facilitates privacy protection, it does not provide a concrete security solution considering the protocol-level challenges. Instead, they suggested major infrastructural changes in 5G settings. 

\noindent \textbf{Concurrent Work.} 
In an independent and concurrent work, \cite{yu_aaka_2024} also uses zero-knowledge proof to address the issue of tracking users' digital footprint in the cellular network. Roughly speaking, \cite{yu_aaka_2024} introduces (AAKA) an AKA protocol that relies on a combination of cryptographic primitives, including Decisional Diffie-Hellman, zero-knowledge proof, BBS signatures, Keyed-Verification Anonymous Credential, and ElGamal Encryption. Despite asserting its compatibility with 5G, including these asymmetric cryptographic elements raises practicality concerns.

The AAKA protocol exhibits several limitations that warrant consideration. Primarily, AAKA is explicitly tailored for complete privacy, Making it hard to revoke a specific user from the network. Additionally, its lack of composability and absence of UC security potentially compromise its robustness when integrated into larger, real-world systems. From a cryptographic perspective, AAKA's reliance on pairing-based settings and the discrete logarithm assumption renders it vulnerable in post-quantum scenarios, as it does not incorporate quantum-secure primitive-based assumptions. Furthermore, While the primary goal of AKAA is to achieve comprehensive privacy, it is important to note that this privacy pertains exclusively to the AKA protocol, leaving the privacy of roaming users during handovers unaddressed. When MVNO users undergo handover, the MNOs also obtain handover data, e.g., information about the source and destination cells. By secure handover (SH), we mean MNOs will be oblivious to this handover. Now, in order to achieve anonymity, particularly at the MVNO setting, one of the possible approaches could be using a thick model and an SM-DP+ that rotates IMSIs. However, this approach cannot prevent linkability attacks and fake base station attacks. In this regard, each time when the IMSI rotates, the attacker can try to relink the victim UE to the new IMSI by replaying the Authentication Request message for the previous IMSI~\cite{basin_formal_2018}. Furthermore, even if the IMSI catcher is not realizable due to the use of public-key encryption of SUPI, the attacker can still launch other fake base station attacks. For example, the attacker can use RRC Reject to launch a DoS attack as described in 5GReasoner\cite{hussain20195greasoner}. While AAKA protocol addresses certain privacy concerns, our proposed protocol offers a more comprehensive and robust solution to the privacy and security challenges in 5G networks. Both approaches improve upon the conventional 5G AKA protocol, which is susceptible to linkability attacks \cite{wang_privacy-preserving_2021}. These attacks exploit the protocol's handling of MAC failures and its predictable challenge-response approach. When a MAC verification fails, the network sends a distinct error message, allowing an attacker to differentiate between a targeted subscriber and others. We achieve unlinkability through a novel mechanism: encrypting all responses from the Core Network using fresh, session-specific public keys for each user. This ensures that even if an adversary forwards a message (e.g., M4) to a group of users, including the intended recipient, none of them can decrypt it. This is because users have already generated new session keys, and without the correct private key, decryption is impossible. Additionally, our protocol's ZKP-based approach guarantees user unlinkability across sessions, as the network never receives user identifiers. As demonstrated in Lemma 2, this comprehensive strategy achieves robust unlinkability throughout the entire protocol, significantly enhancing user privacy and resistance to tracking attacks.

Another key distinguishing feature of our protocol is its resilience against fake base station attacks, an area where previous protocols like AAKA and 5G AKA show limitations. We implement a dual-layered strategy: first, a public key infrastructure for base stations, requiring sanitizable signatures with CN and gNB public keys, dramatically increases the difficulty of simulating legitimate nodes. Second, we maintain comprehensive privacy through ZKP, allowing user authentication without revealing specific information to gNBs. In contrast, conventional 5G-AKA and AAKA remain vulnerable to fake base station attacks due to unauthenticated System Information Block messages \cite{hussain20195greasoner,basin_formal_2018}. They also struggle with privacy-preserving handovers, as MNOs can track user locations through gNB observations. These systems would require impractical UE re-registration with new IMSIs at each base station connection.
Table \ref{tab:sec.Features} offers a comparative analysis of related works, including the AAKA protocol, alongside our proposed scheme. This comparison illustrates the comprehensive nature of our solution in addressing various security and privacy challenges in 5G networks.

\section{Evaluation}
\label{sec:implementation-evaluation-comparison}

We first provide the details of our testbed setup. We also present the performance analysis of the proposed scheme and compare it with the state-of-the-art protocols presented \cite{3rd_generation_partnership_project_3gpp_security_2020,wang_privacy-preserving_2021,fan_rehand_2020,zhang_robust_2021,cao_cppha_2021}.

\begin{figure}
  \centering
  \begin{subfigure}{0.35\textwidth}
    \centering
    \includegraphics[width=1\linewidth]{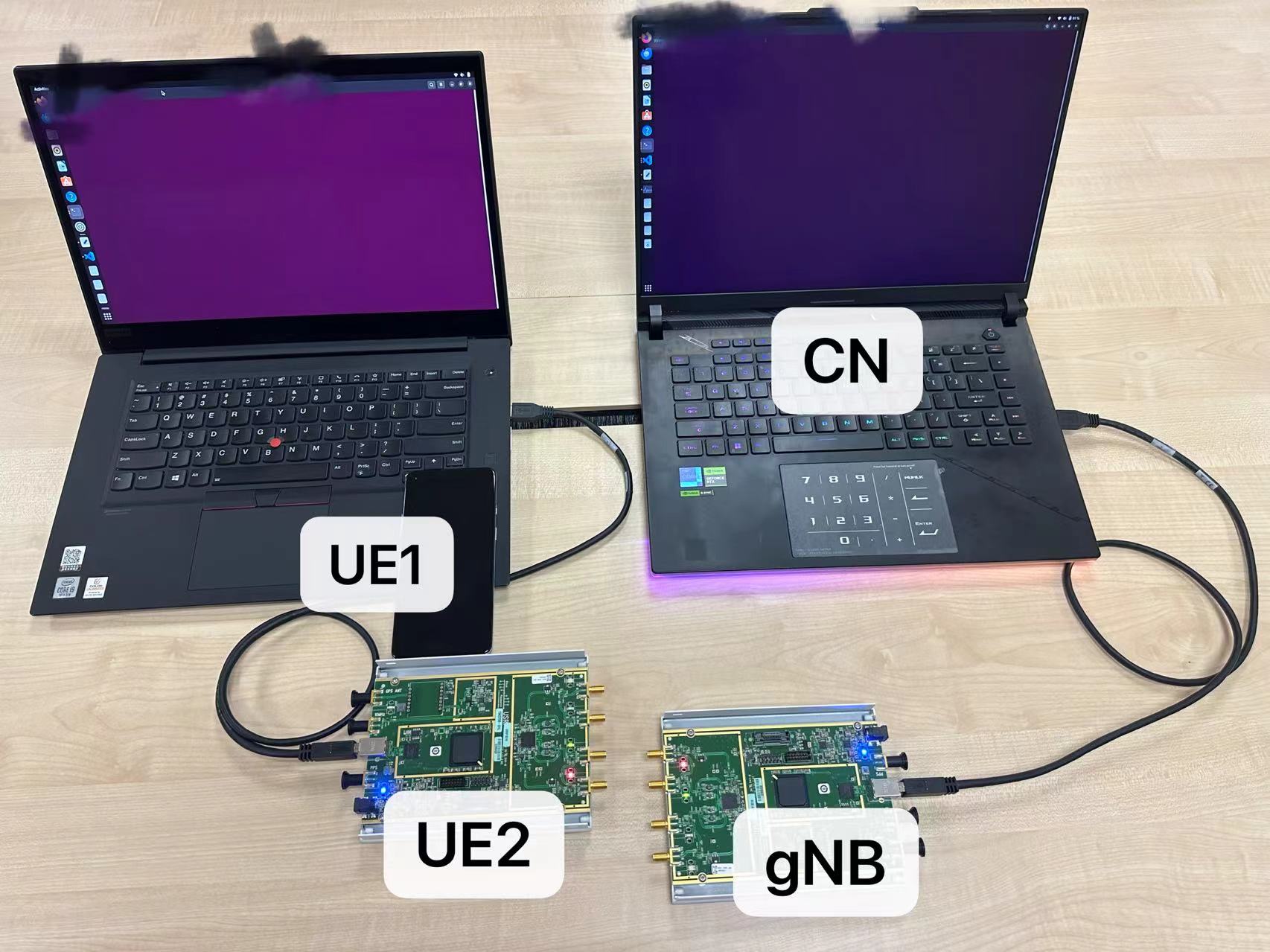}
    \label{fig:sub1}
  \end{subfigure}
  
  \begin{subfigure}{0.45\textwidth}
    \centering
    \includegraphics[width=.95\linewidth]{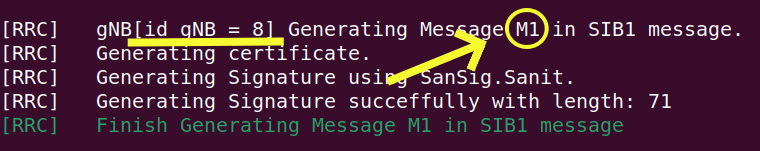}
    \label{fig:sub2}
  \end{subfigure}
  \vspace{-0.3cm}
  \caption{(Top) Testbed Setup, and (Bottom) Log From gNB.}
  \label{fig:testbed}
  \vspace{-0.35cm}
\end{figure}

\subsection{Testbed Setup}

As shown in Figure \ref{fig:testbed}, to build the testbed environment, we use an ASUS machine with an i9 core, 5.6GHz CPU, and 16.0 GB RAM, two USRP B210 \cite{EttusResearch} software-defined radios (SDRs) connected to the computer running an Ubuntu 22.02 desktop OS. We use a popular open-source 5G stack called OpenAirInterface \cite{OpenAirInterface} to set up the UE, 5G base station (gNB) and core network. To implement our protocol on the standard 5G protocol, we modify the code of the core network, gNB, and UE stack in OpenAirInterface to support our initial authentication protocol. In this setup (as shown in Figure  \ref{fig:testbed}), one USRP connected with a Lenovo laptop acts as the 5G base station, and another USRP connected with the second Lenovo Laptop works as a UE ($UE_2$). Both machines run Ubuntu 22.02 operating system on a core i7 core machine with a 2.7GHz CPU and 32.0 GB RAM.

In addition, to effectively measure the performance of our proposed scheme on 5G phones, we use a Galaxy Note 9 smartphone ($UE_1$) running Android 10 mobile operating system and equipped with octa-core processors (1.8GHz Quad-Core ARM Cortex-A55, 2.7GHz Quad-Core Mongoose M3) and 6GB of RAM. The cryptographic operations of our proposed scheme and related works were implemented using OpenSSL 3.0 \cite{openssl}, Java Pairing-Based Cryptography (JPBC) \cite{de_caro_jpbc_2011} and Java Cryptography Extension (JCE) \cite{technology_network_java_2022}. Since we cannot modify the existing firmware of the basebands, we implemented the cryptographic algorithms on the application processor. We measured the computation time to demonstrate the feasibility of the proposed cryptographic constructs on the commodity phones. To comprehensively analyse and compare our proposed scheme against existing works, we implemented the cryptographic algorithms/protocols proposed in the compared papers and measured those works' computational and communication costs (wrt. time) in our testbed. For analysing the communication costs, the communication setup involves considering factors such as propagation and transmission time, message size, and the network's data rate to measure transmission delays across all protocols accurately.

Due to the page limit, we provide only a single log picture of our 5G implementation in Figure \ref{fig:testbed}. This log shows the process of gNB generating message $M_1$ at the initial authentication protocol. Full implementation details can be found in the Appendix. We also provide the prototype of the implementation through GitHub~\cite{github_repo}.

\subsection{Evaluation Results with Testbed}

We have implemented the entire proposed AKA protocol in our 5G testbed. We evaluate it using three metrics. First, we compute the computational cost, i.e., the time required by UE, gNB, and CN to perform the cryptographic operations involved in our proposed protocol. Second, we measure the communication costs, i.e., the number of additional bytes introduced by our approach to the existing 5G-AKA over-the-air messages. Finally, we evaluate the end-to-end latency of the entire AKA protocol. This includes all the computational and communication costs. We also use the same SDR-based testbed to compare the end-to-end latency of our protocol with those of state-of-the-art protocols  \cite{3rd_generation_partnership_project_3gpp_security_2020, wang_privacy-preserving_2021, schmitt2021pretty, yu_aaka_2024}.

\begin{table}[h]
    \caption{Time Required for Cryptographic Operations and Communication Overhead.}
    \centering
    \vspace{-0.3cm}
    \scalebox{0.9}{
    \begin{tabular}{c | c |c | c | c }
    
        \toprule[1.3pt]

        \diagbox[width=8em]{\textbf{Compare}}{\textbf{Message}} & \textbf{M1} & \textbf{M2} & \textbf{M3} & \textbf{M4} \\

        \hline

        \textbf{Time Required (ms)} & 0.416 & 12.236 & 4.876  & 1.679   \\
        \hline

        \textbf{Message Size (bytes)} & 95 & 830 & 830 & 256   \\

    \midrule[1.5pt]
    
    \end{tabular}
    }
    \label{tab:performance compariion SDR-based}
    
    \vspace{-5mm}

\end{table}

\begin{table*}[h!]
    
    \caption{End-to-End Performance comparison with SDR-based Testbed.}
    \vspace{-0.3cm}
    \centering
    \scalebox{0.85}{
    \begin{tabular}{c | c |c | c | c | c }
        \toprule[1.3pt]

        \diagbox[width=8em]{\textbf{Compare}}{\textbf{Schemes}} & \textbf{Conventional-5G \cite{3rd_generation_partnership_project_3gpp_security_2020}} & \textbf{5G-AKA' \cite{wang_privacy-preserving_2021}} & \textbf{PGPP \cite{schmitt2021pretty}} & \textbf{AAKA \cite{yu_aaka_2024}} & \textbf{Ours Scheme}  \\

        \hline

        \textbf{Protocol Type} & 5G & 5G & MVNO & MNO & MVNO*  \\
        \hline

        \textbf{End-to-End AKA Cost (s)} & 1.31 & 1.33 & - & 1.44 + $\kappa$  &  1.41  \\

    \midrule[1.5pt]
    
    \end{tabular}
    }
    \label{tab:performance compariion SDR-based}
    
    \vspace{-2mm}
\end{table*}

\subsubsection{Time Required for Cryptographic Operations}

In this experiment, we report the time required for different entities to run all cryptographic operations for message $M_1$ to message $M_4$ by the corresponding entities. 
In our proposed AKA protocol, the message $M_1$ is generated and sent from the gNB. The major cryptographic operation involved is the generation of the signature, which takes 0.416ms. The message $M_2$ is generated by the MVNO user, and it consists of a verification process, zero-knowledge proof, and a singing phase. Note that proof can be generated offline; hence, we have not considered its associated time. So, the time required to generate $M_2$ is 12.236ms. For message $M_3$, it needs a verification phase and a signing phase, which is around 4.876ms. The last message $M_4$ involves the verification and the signing phase, where the verification of the proof takes longer time, around 1.679 ms.

\subsubsection{Communication Overhead}
Here, we provide the details of each message regarding the overhead. While the message $M_1$ contains the certificate (24 bytes) and its signature (71 bytes), the SIB1 message of our protocol is a total of 95 bytes. In messages $M_2$ and $M_3$, we compute a zero-knowledge proof (304 bytes), a serialised public key (426 bytes), a timestamp (8 bytes) and the signature (71 bytes), which is 830 bytes in total with other supported bytes. In the last message, $M_4$, we send encrypted data using a public encryption scheme, which is 256 bytes.

\subsubsection{\textbf{End-to-End Cost}}
The goal of this experiment is to measure the overall latency of the protocol. To better illustrate the results, we measure the time required by our proposed AKA scheme and other state-of-the-art protocols from the time that UE receives the SystemInformationBlockType1 (SIB1) (i.e., M1 in Figure~\ref{fig:5G-AKA}) message to UE sends the authentication response (Next message of M4 in Figure~\ref{fig:5G-AKA}).  It includes the computational cost (cryptography part), communication cost (transmission part), and all other delays between the UE and gNB. Table \ref{tab:performance compariion SDR-based} provides the results of the state-of-the-art protocols \cite{3rd_generation_partnership_project_3gpp_security_2020, wang_privacy-preserving_2021, schmitt2021pretty, yu_aaka_2024} for the end-to-end cost based on SDR-based testbed. It details the overall latency (time) required for the execution of the initial Authentication (AKA), i.e., AKA protocols, measuring the total time needed to perform the protocol at the User Equipment level ($T_{\UE}$) and the system level ($T_{\text{Sys}}$) (i.e., gNB and CN). As we can see, the conventional 5G protocol takes around 1.32 seconds, which is the baseline of the state-of-the-art protocols. The 5G-AKA' takes a similar amount of time; hence, it changed a small part of the protocol. The AAKA scheme takes a longer time (1.44 + $\kappa$ seconds) than ours because it uses zero-knowledge proof with paring operations and ECDH, which have higher computational costs. $\kappa$ refers to the transmission time needed after UE sends the authentication response. For ours, it takes 1.41 seconds while we are using zero-knowledge proof but we can move the computational overhead out of the real-time phase. Although our proposed scheme exhibits slightly higher latency compared to conventional 5G protocols, it significantly enhances security guarantees relative to existing alternatives. It is important to note that all related works have been compared except for PGPP~\cite{schmitt2021pretty}. Empirical evaluation with PGPP is difficult because PGPP did not make its implementation open-source. Moreover, PGPP requires a dedicated HW/infrastructure, such as a PGPP gateway, which is impossible to implement as the implementation details are missing in their paper. We first introduce the results of the SDR-based testbed.

\begin{table}[h!]
    \caption{Performance comparison of only the cryptographic operations on the application processor of a commercial phone (Galaxy Note 9).}
    \vspace{-0.3cm}
    \centering
    \scalebox{0.85}{
    \begin{tabular}{c |c |c c }
        \toprule[1.3pt]


        \multirow{2}{*}{\textbf{Schemes}} &\multirow{2}{*}{\textbf{Type}} & \multirow{2}{*}{\textbf{Phase}} & \multirow{2}{*}{\textbf{ $T_{Phone}$ (ms)}} \\
        &  &  &  \\
        
        \bottomrule[1.3pt]

        \multirow{3}{*}{Conventional-5G \cite{3rd_generation_partnership_project_3gpp_security_2020}}& \multirow{3}{*}{5G} & IA& 2.82
        \\  \cline{3-4}
        &&HO  &2.7
        \\ \cline{3-4}
        && Total (IA+HO) &5.52
        \\
        \hline

        \multirow{3}{*}{\multirow{1}{*}{$5G-AKA' $ \cite{wang_privacy-preserving_2021}}}& \multirow{3}{*}{5G} & IA& 1.69
        \\  \cline{3-4}
        &&HO  &1.99
        \\ \cline{3-4}
        && Total (IA+HO) &3.68
        \\
        \hline

        \multirow{3}{*}{PGPP \cite{schmitt2021pretty}} & \multirow{3}{*}{MVNO}& IA&-
        \\  \cline{3-4}
        &&HO &NS
        \\ \cline{3-4}
        && Total (IA+HO) &-
        \\
        \hline

        \multirow{3}{*}{AAKA \cite{yu_aaka_2024}} & \multirow{3}{*}{MNO}& IA &108.95
        \\  \cline{3-4}
        &&HO &NS
        \\ \cline{3-4}
        && Total (IA+HO) &-
        \\
        \hline

        \multirow{3}{*}{Ours} & \multirow{3}{*}{MVNO*} &IA&5.65\\  
        \cline{3-4}
        &&HO&5.45
         \\ 
         \cline{3-4}
        && Total(IA+HO) &11.11
        \\

    \midrule[1.5pt]
    
    \end{tabular}
    }
    \label{tab:performance compariion Android}
    
    \vspace{-6mm}

\end{table}


\subsection{\textbf{Evaluation of Cryptographic Operations on Phone and without Testbed}}

We also evaluate the time required by a commercia-gradel UE to perform the cryptographic operations involved in different AKA and Handover protocols. For this, we use a Galaxy Note 9 as the UE and implemented only the cryptographic operations (e.g., signature and zero-knowledge proof generation and verification) of the Authentication (AKA) and Handover (HO) protocols without implementing the message flows between UE, gNB, and CN. We  compute the time required by Galaxy Note 9 to perform those cryptographic operations during the executions of different AKA and Handover protocols. Note that it is difficult to modify the baseband to run the protocols, so we use the application processor of Galaxy Note 9 to perform the experiment, which is reasonable for the purpose of comparing the relative difference between our protocol and the other state-of-the-art protocols.

Table~\ref{tab:performance compariion Android} presents the computation time required for the execution of both the initial authentication (AKA) and Handover (HO) protocols, measuring the total time needed to perform the protocol at the phone level ($T_{phone}$). It shows that our proposed initial authentication protocol requires approximately 5.65 ${\mu s}$.  On the other hand, our proposed Handover protocol requires approximately 5.45 ${\mu s}$. Compared with the AAKA protocol, our scheme takes less time for the initial authentication phase. Meanwhile, compared with the conventional 5G protocol, even though our proposed protocol takes longer time, our scheme provides higher security guarantees.

\section{DISCUSSION}

    \noindent$\bullet$ \textbf{Lawful traceability}: In general, to ensure comprehensive privacy, we do not allow MNOs to access the user's identity. Thus, in our current solution, the MVNO needs to incorporate MNOs to locate individuals legally. With this aim, we propose a new definition of revocation, discussed in Section~\ref{Efficient Revocation}, alongside a practical ZKP, where the MVNO generates the setup for the ZKP. The combination of these elements allows us to maintain user privacy by using non-identifying yet unique attributes to trigger revocation. However, in cases requiring lawful traceability, the MVNO is obligated to share secret information with MNO, including the trapdoor of $crs$. This approach ensures compliance with legal requirements without compromising user identities.

    \noindent$\bullet$ \textbf{Lawful revocation}: The revocation procedure of our proposed privacy-preserving AKA protocol is crucial due to the protocol's strong emphasis on user anonymity and unlinkability. This anonymity can complicate identifying and revoking a specific user's credentials, which is essential for handling fraudulent activities or compromised user data. Traditionally, revocation mechanisms rely on identifying information to disable an account, but this approach conflicts with the privacy guarantees of our protocol. To address this, we propose a new definition of efficient revocation, explained in Section~\ref{Efficient Revocation}, that allows for effectively managing network security and legal compliance without revealing user identities. This mechanism is crucial for handling instances of fraud or compromised data while keeping to stringent data protection laws like GDPR in Europe, which require the ability to disable access for users engaging in illegal activities without breaching privacy regulations.

\vspace{-3mm}

\section{Related Work}

\label{sec: related work}
Considerable efforts have been dedicated to enhancing the security of 5G networks, particularly in the realm of user authentication and network security during authentication and handover protocols. Here, we highlight some under-specified security and privacy requirements and weaknesses in the current 5G-AKA and HO protocol versions.  These issues were previously addressed by \cite{cremers_component-based_2019,braeken_symmetric_2020,basin_formal_2018, schmitt2021pretty}, which include vulnerabilities such as traceability attacks from active adversaries, identity confusion attacks, lack of perfect forward secrecy, and confidentiality attacks on sequence numbers. In response to these identified vulnerabilities, ongoing efforts have been undertaken to develop effective countermeasures \cite{braeken_symmetric_2020, fan_rehand_2020, zhang_robust_2021, wang_privacy-preserving_2021, singla_look_2021}.
Addressing user privacy concerns is a pivotal challenge within the 5G-AKA framework, an issue that is thoroughly examined by Braeken \cite{braeken_symmetric_2020} and Wang et al.\cite{wang_privacy-preserving_2021}. To tackle the anonymity and unlinkability issues inherent in the current version, they have introduced an improved 5G-AKA protocol. Beyond privacy, the assurance of base station authenticity in user communication is another critical issue in preventing potential fake base station attacks. In this regard, a robust solution has been proposed in \cite{singla_look_2021}, presenting an efficient approach to mitigating this particular security challenge. In the context of the 5G Handover (5G-HO) protocol, \cite{fan_rehand_2020} introduces a region-based handover protocol (ReHand) that ensures user anonymity, perfect forward secrecy, and a fast revocation mechanism.
Similarly, \cite{zhang_robust_2021} have designed a universal HO protocol using chameleon hash functions and blockchains (RUSH). This protocol achieves security features similar to those of the ReHand protocol. However, it is essential to note that both protocols (Rehand and RUSH) rely on the standardized 5G-AKA protocol, which is vulnerable to perfect forward secrecy and only supports weak anonymity \cite{basin_formal_2018}. As a result, the security of their proposed protocols is affected by the security and privacy weaknesses of the current version of 5G-AKA. Despite the extensive efforts in this domain, the current works in the field fall short of accomplishing privacy-preserving and secure authentication and handover protocols. 
Furthermore, none of the prior works investigated or presented the critical requirement of achieving universal composability in the context of security protocols. Additionally, the security and privacy aspects specific to the 5G-enabled MVNO environment remain unexplored in prior research, rendering this work the first, to the best of our knowledge, in this particular area.

\vspace{-3mm}

\section{Conclusion and Future Work}
In recent times, the Mobile Virtual Network Operator (MVNO) has garnered significant interest. MVNOs provide a multitude of advantages that render them an attractive option for the majority of consumers. In this paper, we first shed light on some of the prominent security and privacy issues in the MVNO environment, where an MVNO needs to share its customer information with the MNO for validation. This may cause serious privacy issues for their customers. In order to address all these issues, here we have introduced a universally composable authentication and handover strategy that provides robust user privacy. The scheme allows any MVNO user to verify a mobile operator (MNO) and vice versa while ensuring user anonymity and unlinkability support.  Our proposed solution is expected to be implemented by the MVNO(s) in order to provide improved privacy support to their customer(s). One of the proposed directions for future work would be to consider cross-layer authentication under MVNO settings.     

\noindent\textbf{Future Work.} This work does not consider privacy leakage through physical layer and cross-layer communications, e.g., interactions between physical and upper layers. We consider it our future work.

\vspace{-3mm}

\section*{Acknowledgements}
We thank the anonymous reviewers and the shepherd for their feedback and suggestions. We also thank the vendors for cooperating with us during the responsible disclosure. The work of Prosanta Gope was supported by The Royal Society Research Grant under grant RGS$\backslash$R1$\backslash$221183. Meanwhile, this work of Syed Rafiul Hussain has been supported by the NSF under grants 2145631, 2215017, and 2226447, the Defense Advanced Research Projects Agency (DARPA) under contract number D22AP00148, and the NSF and Office of the Under Secretary of Defense-- Research and Engineering, ITE 2326898, as part of the NSF Convergence Accelerator Track G: Securely Operating Through 5G Infrastructure Program.

\bibliographystyle{splncs04}
\bibliography{mybibliography2,moreRef, 5g}

\section*{Appendix A: Implementation Details}

\begin{figure}[htp]
    \centering
    \includegraphics[width=.9\linewidth]{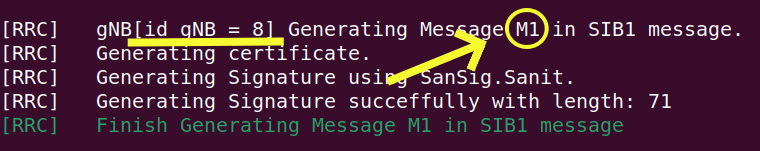}
    \caption{Log From gNB Generating Message 1.}
    \label{fig:app_1}
\end{figure}

Here, we provide the implementation details. Based on the testbed setup we mentioned in Section 7, all results here are generated in an \textbf{openairinterface environment}. We provide the log screenshot using the sequence of the protocol. We first provide the generation of message M1. As shown in Figure \ref{fig:app_1}, gNB generates M1 in the SystemInformationBlockType1 (SIB1) message and broadcasts it.

\vspace{-2mm}

\begin{figure}[htp]
  \centering
  \begin{subfigure}{0.5\textwidth}
    \centering
    \includegraphics[width=.75\linewidth]{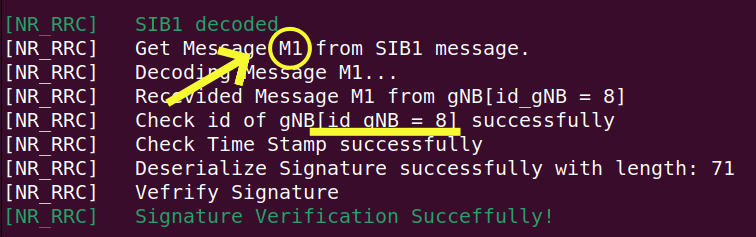}
    \caption{UE Verify M1}
    \label{fig:ue11}
  \end{subfigure}
  \begin{subfigure}{0.5\textwidth}
    \centering
    \includegraphics[width=.75\linewidth]{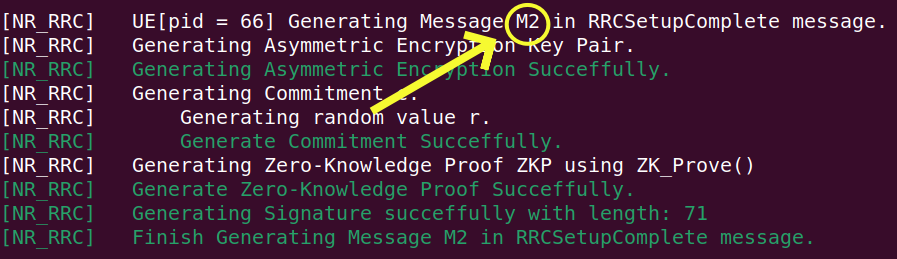}
    \caption{UE Generate M2}
    \label{fig:ue12}
  \end{subfigure}
  \caption{(a) UE Verify M1, and (a) UE Generate M2.}
  \label{fig:ue1}
\end{figure}

After receiving the broadcast message, UE can authenticate the gNB before connecting to it, which is shown in Figure \ref{fig:ue11}. After that, UE generates the message M2 in the RRCSetupComplete message, as shown in Figure \ref{fig:ue12}.

\begin{figure}[htp]
    \centering
    \includegraphics[width=.85\linewidth]{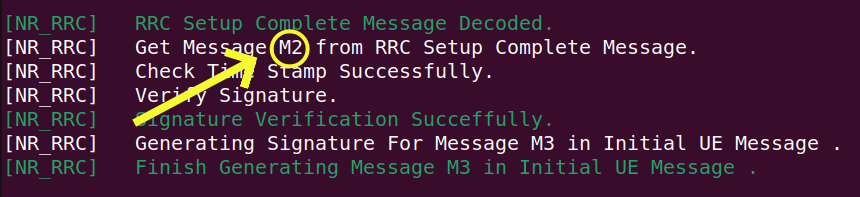}
    \caption{Log From gNB to Generate M3.}
    \label{fig:app_4}
\end{figure}

Upon receiving message M2, gNB first checks the signature from UE and generates the message M3, which is shown in Figure \ref{fig:app_4}. The new message merged into the initial UE message and sent to the core network.

\begin{figure}[htp]
    \centering
    \includegraphics[width=.80\linewidth]{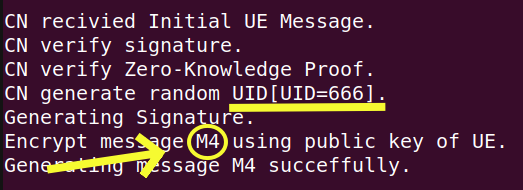}
    \caption{Log From Core Network (CN) to Generate M4.}
    \label{fig:app_5}
\end{figure}

\vspace{-1mm}
As shown in Figure \ref{fig:app_5}, after receiving message M3 from gNB, the core network verifies the gNB signature and the zero-knowledge proof from UE. While the verification is successful, the core network generates the UID, encrypts it using the public key of the UE and sends the message M4 via gNB to UE.

\begin{figure}[htp]
    \centering
    \includegraphics[width=.85\linewidth]{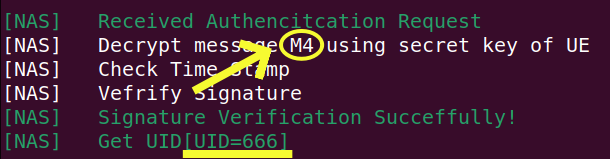}
    \caption{Log From UE to Verify M4.}
    \label{fig:app_6}
\end{figure}

\vspace{-1mm}
Figure \ref{fig:app_6} shows the last step of our protocol. Upon receiving message M4, UE decrypts the message using its own secret key. After verification of the signature of the core network, the UID is retrieved from the core network, and the initial authentication process is finalized.

\vspace{-2mm}

\section*{Appendix B: UC VS Dolev-Yao Adversary 
Model}

 Here are more detailed reasons for the benefits of the UC model over the Dolev-Yao model:

\textbf{1. Robust Composability:} in the Universal Composability (UC) framework ensures that a security protocol remains secure when combined with other protocols, even in complex, dynamic environments. This feature is crucial because real-world applications often involve multiple interacting protocols, each potentially influencing the others' security properties.
The Dolev-Yao model, however, typically examines protocols in isolation. It assumes that the cryptographic primitives are secure and focuses on what happens within a single protocol without external interactions. This isolation can miss critical vulnerabilities that only manifest when protocols operate concurrently or are integrated within larger systems. Thus, while useful for initial protocol design and understanding fundamental vulnerabilities, Dolev-Yao might not adequately predict security failures in more interconnected and realistic scenarios.

\textbf{2. Adversarial Flexibility:} In the UC model, the adversarial capabilities are modeled more comprehensively, including the ability to adapt based on observed interactions, which more closely mirrors potential real-world attacks. The Dolev-Yao model, while useful for basic protocol analysis, operates under more constrained assumptions about adversary capabilities. More precisely, Universal Composability (UC) excels because it simulates adversarial behaviour in a way that mirrors potential real-world tactics. This includes adaptive strategies where the adversary can dynamically adjust their actions based on information gained during protocol execution. It's designed to handle unexpected interactions and coordinated attacks, maintaining security even under such complex conditions. Dolev-Yao, however, assumes a static adversary bound by initial assumptions about capabilities and unable to adapt to changing circumstances during protocol execution. This model does not account for adversaries learning and adapting, which can leave analyzed protocols vulnerable to more sophisticated, real-world attacks that exploit dynamic or unexpected conditions.

This difference makes UC more suitable for modern cryptographic systems where adaptability and resilience against sophisticated threats are crucial.

\textbf{3. Ideal vs. Real World Simulation:} UC security proofs involve demonstrating that no adversary can distinguish between the ideal functionality (a theoretical, perfectly secure system) and the real protocol implementation. This method provides a high level of assurance that all aspects of protocol security (secrecy, integrity, authentication) are preserved under any operational context, a perspective less emphasized in the Dolev-Yao model.
    
More precisely, The Dolev-Yao model's approach to adversarial flexibility is limited because it fundamentally views adversaries through a symbolic lens, assuming they can manipulate and intercept communications but cannot break well-established cryptographic primitives. This model does not account for more nuanced, real-world adversarial behaviours such as side-channel attacks, stateful attacks, or dynamic responses to changing protocol states. In practical scenarios, these limitations can hinder the model's ability to fully predict and counter sophisticated or adaptive threats that might exploit specific implementation flaws or emerging vulnerabilities. The UC model, in contrast, allows for a broader range of adversarial capabilities and interactions, reflecting more realistic and complex attack scenarios. However, the Universal Composability (UC) model's strength in adversarial flexibility arises from its comprehensive approach to modelling potential attacker behaviours. Unlike more static models, UC allows for adaptive adversaries who can change their strategies based on observed interactions and outcomes within the protocol execution environment. This dynamic capability is essential for assessing security in realistic scenarios where threats evolve and where interactions between different components can lead to unforeseen vulnerabilities. The UC model's ability to handle such complex, interactive situations makes it particularly robust and suitable for modern cryptographic applications where security needs to be assured even under varied and potentially hostile operational conditions.

\end{document}